\documentclass[twocolumn,prl,aps,showpacs,floatfix,superscriptaddress]{revtex4}
\usepackage{epsfig}
\usepackage{graphicx}
\begin{document}

\widetext
\title{Short-time scales in the Kramers problem: past, present, future 

(review and roadmap dedicated to the 95th birthday of Emmanuel Rashba)}

\author{Stanislav M.\ Soskin}
\email{stanislav.soskin@gmail.com}
\affiliation{Institute of Semiconductor Physics, National
Academy of Sciences of Ukraine, 03680 Kyiv, Ukraine}
\author{Tetiana L.\ Linnik}
\email{linnik1971@hotmail.com}
\affiliation{Institute of Semiconductor Physics, National
	Academy of Sciences of Ukraine, 03680 Kyiv, Ukraine}
\affiliation{Experimentelle Physik 2, Technische Universität Dortmund, 44227 Dortmund, Germany}

\date{\today}

\begin{abstract}
The problem of noise-induced transitions is often associated with 
Hendrik
{\it Kramers} due to his seminal paper of 1940, where an archetypal example - one-dimensional potential system subject to linear damping and weak white noise - was considered and the quasi-stationary rate of escape over a potential barrier was 
estimated for the ranges of extremely small and moderate-to-large damping. The gap between these ranges was covered in the 80th by 
one of Rashba's favourite disciples Vladimir Ivanovich 
Mel'nikov. 

It is natural to pose a question: {\it how does the escape 
rate 
achieve the quasi-stationary stage}? At least in case of a single potential barrier, the answer seems to be obvious: the escape 
rate
should smoothly and monotonously grow from zero at the initial instant to the quasi-stationary value at time-scales of the order of the time required for the formation of the quasi-stationary distribution within the potential well. Such 
answer
appeared to be confirmed by the analytic 
work of
Vitaly
Shneidman in 1997. However our works in the end of the 90th and in the beginning of the 2000th in collaboration with one more Rashba's favorite disciple Valentin Ivanovich Sheka and with Riccardo Mannella showed that, at a shorter time-scale, namely that of the order of the period of natural oscillations in the potential well, the escape 
rate
growth generically occured {\it stepwise} or even in an oscillatory manner. Analytic results were confirmed with computer simulations. 

In the {\it present paper}, we review those results and provide a roadmap for the development of the subject, in particular demonstrating that various recently exploited experimental systems are excellent candidates for the observation of the above non-trivial theoretical predictions and, moreover, they 
promise
useful
applications.  
\end{abstract}

\maketitle

\section{I. Rashba's influence on our lives}

Prior to the passing to a purely scientific part of the paper, we would like to do an informal introduction dedicated to Emmanuil Iosifovich Rashba (to whom, for the sake of brevity, we further refer as EIR) and to tell about 
his role in our life.
SMS tells as follows.

\lq\lq 
I have been knowing EIR for about 57 years.
In 1965, when I was about five-years old, our family moved to a house 103/3 at Bolshaya Kitaevskaya street in Kiev. EIR  lived in a neighbouring section of the same house and I 
met
him sometimes 
in the yard.
My father 
said to me:
\lq\lq
This man is an outstanding physicist-theoretician!''
I was however 
more interested in his daughter Yulia. 
Being 
about
my age,
she was beautiful, smart, and amiable. I liked playing with her in the yard of our house. 
Regrettably for me, the communication with Yulia stopped rather soon: 
in 1966,
her father accepted an offer to 
head
the Theory of Semiconductors Division at the recently founded Institute of Theoretical Physics in Chernogolovka
and moved there together with the
family.

My next (implicit) intersection with EIR occured as a meeting with his scientific \lq\lq child'', namely one of his favourite disciples Vladimir Ivanovich Mel'nikov, who followed the mentor in his moving from Kiev to Chernogolovka. I was introduced to Mel'nikov by my mentor Mark Isaakovich Dykman in 1986 in Moscow at the General Meeting of the Academy of Sciences of USSR devoted to the discovery of the high-temperature superconductivity. My mentor asked Mel'nikov to be the primary referee of my thesis for the candidate of sciences degree (the PhD analogue). Mel'nikov agreed. 
Allowing for
that, I decided to study his papers in order to properly cite them in the thesis. I found then that he had done an extraordinarily beautiful 
work \cite{Melnikov:84,Melnikov_Meshkov:86} (see \cite{Melnikov:91} for review and more references) on the matching between strongly underdamped and moderate-to-large-damping limits in the Kramers problem of the quasi-stationary escape from a potential well induced with a weak white noise \cite{Kramers:40} (this problem laid unsolved for more than 40 years despite numerous attempts of its solution \cite{Melnikov:91}). 

Although I never used 
immediate results of Mel'nikov's work, it played 
an
important role in my scientific life. Indeed, I used one of the auxiliary ideas exploited by Mel'nikov in his work, namely the idea to use a Gaussian distribution with an increasing in time width in certain integral equation. This helped me to obtain a semi-explicit solution of a problem comletely different from the Kramers problem, namely to find a universal shape of characteristic peaks in fluctuation spectra of a broad class of systems, which inherently required to find a non-trivial dynamics of a system in contrast to the quasi-stationary distributions considered by Mel'nikov. When I showed the semi-explicit results to my mentor, he suggested that these results indicated that, the corresponding non-stationary Fokker-Planck equation (FPE) for the probability density could be reduced to some universal equation and might be it would be possible to solve it. His keen intuition turned out 
right: in the relevant underdamped asymptotic limit, I reduced the complicated general form of the FPE to a relatively simple but still non-trivial equation in partial derivatives of the 2nd order and, even more important, managed to find its explicit solution, which allowed me in turn to find the universal shape of spectral peaks in such a class of systems. I consider this work \cite{Soskin:89} as one of my best ones. Moreover, 
it have played
the major role in an identification of certain variety of fluctuational and dynamical phenomena as a characteristic class of the so called zero-dispersion phenomena (see \cite{Soskin:03,Soskin:12,Huang:19} and references therein). 

Not only was I 
glad to have solved
the important physical problem in \cite{Soskin:89} but I was also proud with the fact that I was seemingly the first person who had solved this non-trivial differential equation in partial derivatives of the 2nd order.
When
I worked on some quantum problem a few years afterwards, I was studying the book \lq\lq Statistical Mechanics'' by Feynman \cite{Feynman:72} and found there $\dots$ almost the same equation
in a completely different physical context! The problem where it arose in studies by Feynman was a problem about the stationary density matrix in a harmonic oscillator: the variables of the properly normalized reciprocal temperature and coordinate in Feynman's problem corresponded accordingly to the variables of the properly normalized time and energy in my problem. The only difference between the normalized equations was that the multiplier of the nontrivial (quadratic) term in Feynman's equation was purely real while that in mine was purely imaginary. Correspondingly, the structure of the solution from the physical point of view strongly differed while the mathematical one was the same. The way in which Feynman obtained the solution was exactly the same as 
I did.
He was apparently so excited with the beauty of this way that, similarly to me, presented it in full detail. 
I was disappointed by the fact that 
the priority in the solution of the mathematical equation was not mine,
but the disappointment was decreased due to a feeling that I had something in common with such an outstanding scientist as Richard Feynman. In order to 
transform
this thrilling story into the closed loop, I need to add the following. Mentors of Mel'nikov and Dykman (whose ideas inspired me, as demonstrated above) were EIR and Mikhail Alexandrovich Krivoglaz respectively while both Rashba and Krivoglaz had one and the same mentor - Solomon Isaakovich Pekar; and the loop is being closed as follows: in the famous book \lq\lq Quantum Mechanics and Path Integrals'' \cite{Feynman:65}, Feynman refers only to 26 sources (a very small amount as for a book, which means that only truly fundamental sources were referred) and 2 of these sources were papers by Pekar! The story on the whole demonstrates amazing links threading the world in terms of science, geography, time (generations), and, in a sense, noosphere.     

My next implicit intersection with EIR occured via collaboration with another his favourite disciple - Valentin Ivanovich Sheka - and Sheka's disciple Tat'yana Leontievna Linnik. The most exciting results of this collaboration will be presented in the next sections. So, I do not go in details here. Rather I just mention that the collaboration lasted since 1999 till 2005 and resulted in 11 papers (most important of which are \cite{Soskin:00_Lakes,Soskin:01_PRL,Soskin:01_PRL2,Soskin:01_FNL,Soskin:01_Chaos,Soskin:05,SPQEO-2022}) and, in addition, stimulated 1 more my paper without coauthors \cite{Soskin:06}. The collaboration could last more but Sheka's health problems and my intensive involvement in other projects had led to its interruption.
 
It turned out that EIR followed my joint activity with Sheka, and we communicated with him a few times via email on this and other occasions. The most active communication was in July 2021, when I organized the ZOOM seminar dedicated to the memory of V.I. Sheka who died of coronavirus on the 7th of February 2021. EIR was a key speaker at the seminar, and I enjoyed both a communication with him on the eve of the seminar and, yet more so, listening to his lecture (lasting almost an hour!) about the development of theoretical physics in Kiev in the 40th-50th of the previous century, the seminal joint paper with Sheka \cite{Rashba:59}, the current state of art and perspectives of spintronics. 

In one of the letters to my father (with whom EIR was in friendly relations for more than 60 years until the very death of my father in 2020), EIR wrote: "The greatest present which one could get in old age is a clear mind." {\it EIR has been lucky to get such a present}.''

TLL tells as follows. 
\lq\lq 
Though I did not immediately work with EIR, I did feel his influence through his disciple Valentin Ivanovich Sheka (VIS) who became my mentor and a very close person for me for the thirty years of our communication with each other. I felt the influence of EIR especially strongly during the work on the book \cite{Sheka:17}. The book was based on the course of lectures presented by VIS at the Physics Department of Kiev State University in the 60th-80th of the previous century and more recently by me. Many works of EIR were immediately used in the book. I was particularly pleased to know about a high evaluation of the book by EIR and, yet more so, to hear in the lecture by EIR at the aforementioned seminar dedicated to the memory of VIS that the book is highly valued by some of Russian-speaking scientists in the USA. In particular, EIR told that Prof. Lev Levitov preferred using just this book for his teaching a similar course in MIT rather than the first book on the subject \cite{Bir:72} because our book was much easier for a perception by students and because we illustrated the efficiency of the method of invariants at a few modern systems and materials, for example graphene. Concluding this personal dedication, I would like to say that I am very happy to have such a \lq\lq scientific grandfather'' as EIR, and I wish him to further keep his love to physics and life on the whole and to inspire younger researchers.''

\section{II. INTRODUCTION to the scientific part}

In his seminal work \cite{Kramers:40}, Kramers considered a weak
noise-induced flux from a single metastable classical potential well,
i.e.~he considered a stochastic system
\begin{eqnarray}
	& & \ddot {q} +\Gamma \dot {q} + dU/dq =f(t)  , \\
	& & \langle f(t) \rangle = 0, \quad \langle
	f(t)f(t^\prime)\rangle =2 \Gamma T \delta (t - t^\prime) , \quad T\ll\Delta U,
	\nonumber
\end{eqnarray}

\noindent which was put initially at the bottom of a metastable
potential well $U(q)$ with a barrier $\Delta U$, and he then calculated
the quasi-stationary probability flux across the barrier. Models of
type (1) are relevant to chemical reactions \cite{Kramers:40}, SQUIDs \cite{likharev}, nano/micro-mechanical resonators (see e.g. \cite{Huang:19,dykman:22} and references therein), nano-particles in optical traps \cite{Flajsmanova:20} and other
real systems 
(see some of the references in \cite{Melnikov:91}).

There have been many developments and generalizations of the Kramers
problem 
but both Kramers and most
of those who followed him considered only the {\it quasi-stationary}
flux, i.e.~the flux established after the formation of a
quasi-equilibrium distribution within the well (up to the barrier). The
quasi-stationary flux is characterized by a slow exponential decay in
time $t$, an Arrhenius dependence on temperature $T$, and a relatively
weak dependence on friction $\Gamma$:
\begin{equation}
	J_{qs}(t)=\alpha_{qs}{\rm e}^{-\alpha_{qs}t}, \quad\quad
	\alpha_{qs}=P {\rm e}^{-\frac{\Delta U}{T}},
\end{equation}
\noindent where $P$ depends on $\Gamma$ and $T$ in a non-activated way.

But how does the flux evolve from its zero value at initial time
to its quasi-stationary regime (2) at time-scales exceeding the
time $t_f$ for the formation of quasi-equilibrium? The answer may
obviously depend on initial conditions and a relevant boundary
(i.e.\ the boundary through which the escape occurs). As for the
boundary, it can be shown that the most general qualitative
features of the flux are valid for any type of boundary (for
the sake of simplicity, we shall consider below only the absorbing
wall). As for the initial conditions, their relevance may vary.
The simplest and often relevant
initial state is the bottom of the well, since it is the stable
stationary state in the absence of noise: if the noise (not
necessarily of the thermal origin) is switched on at some instant,
then the time evolution of the escape from the {\it bottom} becomes
relevant. It should be emphasized however that, if the relevant
metastable part of the potential is multi-well, then the flux
during the major part of the relevant time is not sensitive to the
initial state provided it is concentrated just in one well (e.g.\
it may be thermalized in the well). As for the single-well case,
the flux evolution is more sensitive to the initial state and we
shall consider various cases. But, first, let us discuss the most
simple case where the initial state is at
the bottom of the potential. We shall refer to it as the {\it bottom initial state}.

It may seem natural to assume that the flux evolution from zero to the
quasi-stationary regime is a monotonic function without any \lq\lq
irregularities''. Apart from the naive argument that \lq\lq noise
smooths everything", this assumption appears sound because the
probability distribution $W$ is distinctly centered at the bottom of
the well both initially and in the quasi-stationary stage: $
W(q,\dot{q}, t=0 |q_0=q_b, \dot{q}_0=0) =\delta(q- q_b)
\delta(\dot{q})$ while $W(q,\dot{q},t\gg t_f |q_0=q_b, \dot{q}_0=0)$ is
a narrow peak of width $\propto \sqrt{T}$ around that same state $\{q=
q_b,\dot{q}=0\}$. Moreover, it was shown in 
Ref. 24
that, both in the underdamped and overdamped limits, the escape flux
$J(t)$ does grow at $t\sim t_f$ in a simple manner.

Despite the above arguments, it can be shown that, generically, $J$
evolves from $J(0)=0$ to $J_{qs}(t\gg t_f)$ in a quite complicated way.

\begin{enumerate}
	
	\item As shown in 
	Sec. III.A, the
	flux grows {\it step-wise} on time-scales of the order of a {\it period
		of eigenoscillation} in the bottom of the well. Apart from a purely theoretical interest in filling the \lq\lq gap'' in time-scales in the Kramers
	problem, this part of our work \cite{Soskin:00_Lakes,Soskin:01_PRL,Soskin:01_PRL2,Soskin:01_FNL,Soskin:01_Chaos,Soskin:05,Soskin:06} was motivated by the
	growing interest in the 
	short time-scales that became relevant in the 90th to
	some
	experiments, 
	e.g.
	those studying chemical reactions down to
	femtosecond time-scales \cite{femto}: the period of eigenoscillations relevant to chemical
	reactions in 
	Ref. 25
	is $\sim 1$--$100$ fs.
	
	\item As shown below in Section III.B, the evolution of the flux on
	longer time-scales in a {\it multi-well} metastable
	potential is also distinctly different from the relatively simple monotonic
	function described in 
	Ref. 24:
	$J$ grows sharply on a
	logarithmic time-scale to a value which is typically very different
	from $J_{qs}(0)$ (typically, exponentially larger) and then evolves to
	$J_{qs}(t)$ during the exponentially long time.
	
\end{enumerate}

\noindent It should be emphasized that the qualitative features of $J(t)$
described above are valid for any reasonable definition of the
flux: e.g.\ the full flux through a boundary or just the
first-passage flux, while the boundary may be a given coordinate,
or a boundary of a basin of attraction, or a boundary of the
vicinity of another attractor, etc.

For illustration, we use the potential

\begin{equation}
	U(q)=q-q^3/3
\end{equation}

\noindent
for the single-well case (Fig.\ 1), and

\begin{equation}
	U(q)=0.06 (q+1.5)^2-\cos(q)
\end{equation}

\noindent
for the multi-well case (Fig.\ 6(a)), with an absorbing wall \cite{footb} at
$q=q_{aw}$ in both cases.
Experimentally, the flux is measured in the following way. The
system is placed at an initial state, after which it follows the
stochastic equation (1) until either the coordinate of the wall,
$q_{aw}$, or the time limit \cite{footc}, $t_{l}$, is reached. It
is then reset to the initial state and everything is repeated.
Once the statistics are deemed adequate, we calculate the flux

\begin{equation}
	J(t)\equiv \frac{1}{N_{\rm reset}}\frac{\Delta N(t)}{\Delta t}
\end{equation}

\noindent where $N_{\rm reset}$ is the overall number of resets,
and $\Delta N(t)$ is the number of resets during the interval
$[t,t+\Delta t]$;  $\Delta t$ is chosen to be much smaller than a
characteristic time over which the flux (5) may change
significantly, but large enough to provide $\Delta N(t)\gg 1$
(roughly, the latter is satisfied provided $\Delta t \gg t_l/
N_{\rm reset}$).

The above experimental definition corresponds to the following
theoretical definition of the flux

\begin{eqnarray}
	&&
	J(t)=\int\int d q_0 d \dot{q}_0 \;
	W_{in}(q_0,\dot{q}_0) J_{q_0,\dot{q}_0} (t),
	\\
	&&
	J_{q_0,\dot{q}_0} (t)=
	\int_0^\infty d \dot{q} \; \dot{q} W(q=q_{aw},\dot{q},t|q_0,
	\dot{q}_0),
	\nonumber
\end{eqnarray}

\noindent where $ W_{in}(q_0,\dot{q}_0) $ is a statistical
distribution of the initial coordinate and velocity and $W$ is the
conditional probability density.

The theoretical approach which we use is the method of {\it optimal
	fluctuation} 
(see e.g. Refs. 10 and 28)
whose details in
application to the present problems are given in the next section.
Results of a verification of theoretical results by computer and analog electronic
simulations are also presented.

\begin{figure}
	\centering
	{\leavevmode\epsfxsize=3.2 in\epsfbox{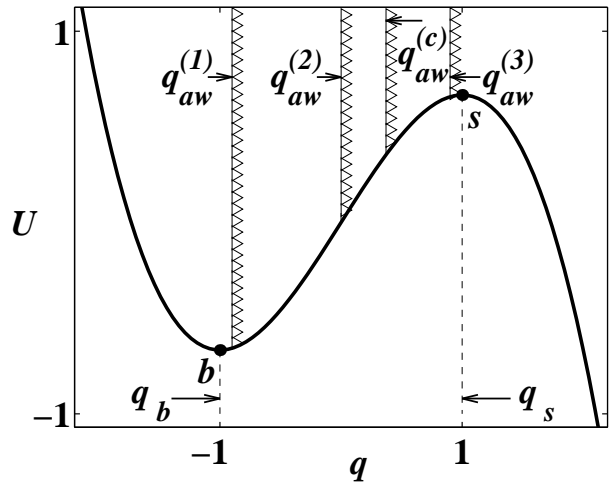}}
	\caption{The potential $U(q)=q-q^3/3$. The bottom and the saddle are
		marked as $b$ and $s$ respectively. Triangles indicate four typical
		positions of the absorbing wall.}
\end{figure}

The structure of the rest of the paper is the following. Sec. III gives the review of former results on the subject. The subsections A and B consider the cases of single-wel and multi-well potentials respectively. In Sec. IV, we 
briefly
present the roadmap for the most interesting developings of the subject including issues of a scientific and practical interest, in the subjects A and B respectively. 
Conclusions are given in Sec. V.

\section{III. Review of former results}

\subsection{A. Single-well metastable potential}
It can be shown directly from the Fokker-Plank equation that the
formation of quasi-equilibrium up to the barrier in the single
metastable well typically takes \cite{formation} a time of the
order of

\begin{equation}
	t_f^{(s)}\sim\frac{1}{{\rm min}(\Gamma,\omega_0^2/\Gamma)}\ln(\frac{\Delta U}{T}),
\end{equation}

\noindent where $\omega_0$ is the frequency of eigenoscillation in the
bottom of the well.

In this section, we shall be interested in much smaller time-scales,

\begin{equation}
	t\ll t_f^{(s)}.
\end{equation}

The work on {\it non-stationary} escape rates in the
Kramers problem preceding to ours was based on the direct solution of the
Fokker-Plank equation (cf.\ \cite{schneidman}).
The
method of {\it optimal fluctuation} to this problem was applied for the first
time 
in \cite{soskin} and then further developed in \cite{Soskin:00_Lakes,Soskin:01_PRL,Soskin:01_PRL2,Soskin:01_FNL,Soskin:01_Chaos,Soskin:05,Soskin:06},
obtaining non-trivial new results for short
time-scales. It is convenient to consider first the case of an
initial state with a {\it given} coordinate and velocity:

\begin{equation}
	W_{in}(q_0,\dot{q}_0) =\delta(q_0-q_i)\delta(\dot{q}_0-\dot{q}_i).
\end{equation}

\noindent
The flux is sought as

\begin{equation}
	J(t) \equiv J_{q_i,\dot{q}_i} (t)=P(t){\rm e}^{-\frac{S_{\rm min}(t)}{T}}
\end{equation}

\noindent where the {\it activation energy} $S_{\rm
	min}(t)$ does not depend on $T$ while the prefactor $P(t)$ depends on
$T$ in a non-activated way. At small $T$ and short $t$, the factor
$\exp(-S_{\rm min}/T)$ depends on $t$ much more strongly than $P$. So,
we concentrate on studying $S_{\rm min}(t)$, which can be shown
\cite{soskin} to be a minimum of the functional:

\begin{eqnarray}
	S_{\rm min}(t) \equiv S_{\rm min}(q_i,\dot{q}_i, t) = {\rm min}_{[q(\tau)],\dot{q}_{aw}} (S), \nonumber\\
	S\equiv S_{\dot{q}(t)}[q(\tau)]=\int_0^{t} d\tau L, \\
	L = (\ddot{q} +\Gamma \dot{q} + dU/dq)^2/(4\Gamma),
	\\
	q(0)=q_i, \quad \dot{q}(0)= \dot{q}_i, \quad q(t)=q_{aw}, \quad \dot{q}(t)=\dot{q}_{aw}.
\end{eqnarray}

\noindent The minimization is done over an escape path $[q(\tau)]$
at a given exit velocity $\dot{q}_{aw}$, with a further minimization
over 
$\dot{q}_{aw}$
\cite{30}. The path
minimizing $S$ may be called the {\it most probable escape path}
(MPEP), in analogy with the quasi-stationary case. The necessary
conditions for the minimum of the functional (11) are as follows.

\begin{enumerate}
	
	\item A zero variation, $\delta S=0$: it implies that the MPEP
	$[q(\tau)]$ satisfies the Euler-Poisson equation
	\cite{elsgolc,soskin}
\begin {equation}
\frac{\partial L}{\partial q} - \frac {d}{dt} (\frac {\partial L}{\partial \dot
	{q}}) + \frac {d^2}{dt^2} (\frac {\partial L}{\partial \ddot {q}}) = 0,
\end{equation}

\noindent which, for the $L$ of the form (12), reads

\begin {equation}
\ddot{} {q}\ddot{} +
\ddot{q}\left(2\frac{d^2U}{dq^2}-\Gamma^2\right) +\dot{q}^2 \frac{d^3U}{dq^3}
+
\frac{d^2U}{dq^2} \frac{dU}{dq} =0.
\end{equation}

\item A zero derivative with respect to the exit velocity, $\partial
S/\partial \dot {q}(t)=0$: this condition can be reduced
to $\partial L/\partial \ddot{q}(t) =0$,
which, for the $L$ of the form (12), reads

\begin {equation}
[\ddot{q} +\Gamma
\dot{q} +dU/dq]|_{\tau=t} =0.
\end{equation}

\end{enumerate}

Solutions of Eq.(15) satisfying three first conditions in (13) and the condition (16) can be found
numerically: in addition to $q(0)$ and $\dot{q}(0)$ given in (13), one
can match $\ddot{q}(0)$ and $\dot{}\ddot{q}(0)$ 
so that the result of
the integration (15) on the interval $[0,t]$ satisfies the third condition in Eq. (13) (i.e. $q(t)=q_{aw}$) and the condition (16).

\begin{figure}
	\centering
	{\leavevmode\epsfxsize=3.2in\epsfbox{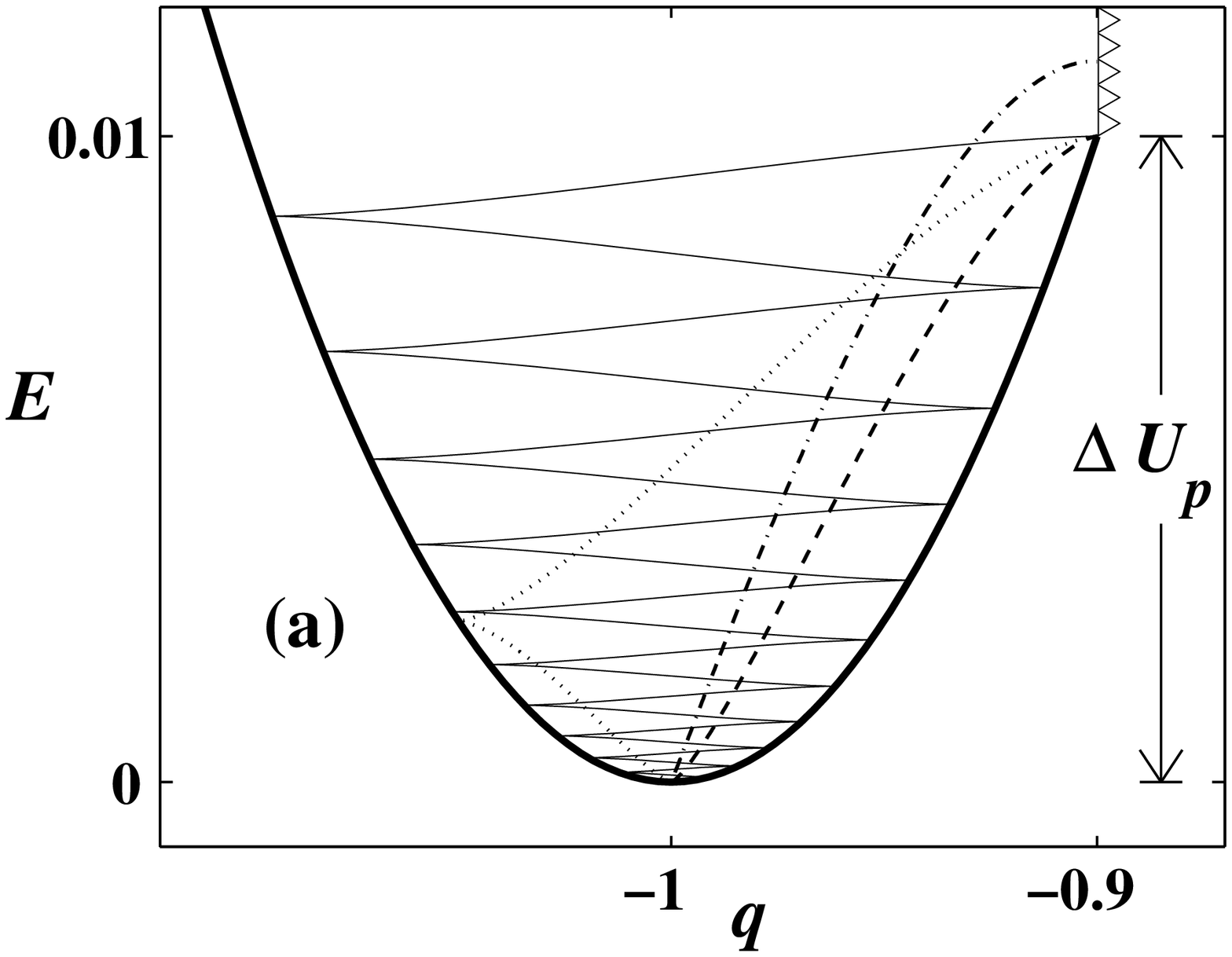}}
	{\leavevmode\epsfxsize=3.2in\epsfbox{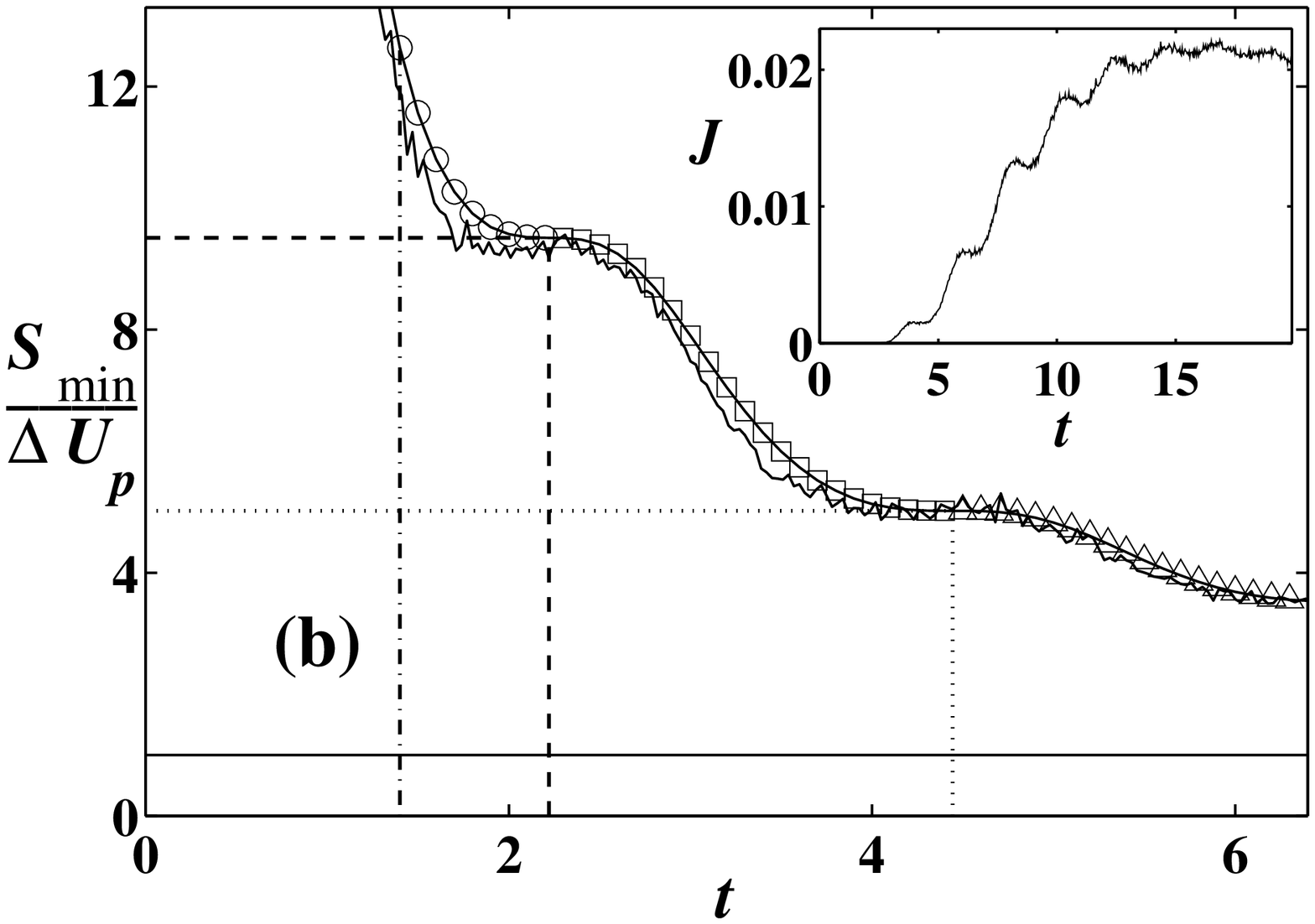}}
	\caption{The case of the bottom initial state. (a) The parabolic
		approximation $U_p(q)\equiv (q+1)^2$ (thick solid line) of
		$U(q)-U(q_b)$ near the bottom, and examples of MPEPs (plotted in the
		energy-coordinate plane $E-q$ where $E\equiv\dot{q}^2/2+U_p(q)$) at
		$\Gamma=0.05$; the absorbing wall (at $q_{aw}=q_{aw}^{(1)}\equiv -0.9$)
		is indicated by triangles; (b) $S_{\rm min}(t)/\Delta U_p$ explicitly
		calculated in the parabolic approximation is shown by the solid line
		with markers: circles, squares and triangles indicate regions
		corresponding to respectively 0, 1 and 2 turning points in the MPEP;
		$S_{\rm min}(t) / \Delta U $ derived from simulations in $U(q)$ (3) is
		shown by the jagged line ($\Delta U \equiv U(q_{aw})-U(q_b)$). Dashed
		and dotted lines indicate the theoretical 1st and 2nd inflection points
		with $dS_{\rm min}/dt=0$, in (b), and the corresponding MPEPs, in (a).
		The thin solid line shows the large-time asymptote level ($=1$), in
		(b), and the corresponding MPEP (which is the time-reversal of the
		noise-free trajectory from the state $(q=q_{aw}, \dot{q}=0)$), in (a).
		The dash-dotted line shows in (a) the MPEP for some arbitrarily chosen
		time $t=1.4$ (see (b)): it demonstrates that the exit velocity is
		typically non-zero. The inset shows $J(t)$ measured at $T= \Delta U$.}
\end{figure}

\subsubsection{1. Bottom initial state }

Let us first consider the case of the bottom initial state: 

\begin {equation}
q_i=q_b,
\quad\quad\quad
\dot{q}_i=0.
\end{equation}

\noindent Before presenting the numerical results, we 
find
important
general features of the MPEPs and $S_{\rm min}(t)$. In particular, we show below that,
as the boundary moves from the close vicinity of the bottom towards the
saddle, $J(t)$ undergoes qualitative changes while still being
step-wise.

First, consider the case when the absorbing wall is close to the
bottom: $U(q)$ may then be approximated by a parabola (Fig.2(a))
\begin{equation}
	U(q)-U(q_b)\approx \frac{\omega_0^2}{2}(q-q_b)^2,
\end{equation}

\noindent where $\omega_0=\sqrt{2}$ and $q_b= -1$, in the case of $U(q)$
(3). Thus (15) reduces to a linear equation with constant
coefficients that can be integrated explicitly. $ S_{\rm min}(t)$ can be
found explicitly too. Rather than presenting 
the
cumbersome formulas,
we discuss their most important consequence: if $\Gamma<2\omega_0$,
then $ S_{\rm min}(t)$ has a step-wise shape (Fig.2(b)) i.e.\ possesses
inflection points with $dS_{\rm min}/dt=0$ at

\begin{eqnarray}
	t=t_n\equiv \frac{n\pi}{\omega_0\sqrt{1-(\Gamma/2\omega_0)^2}}, \quad\quad\quad\quad\quad\quad\quad\quad\quad\quad\nonumber\\
	S(t_n)=\frac{\Delta U_p}{1-\exp(-\Gamma t_n)},\quad\quad\quad\quad\quad\quad\quad\quad\quad\quad\quad\quad\\
	\Delta U_p\equiv \omega_0^2(q_{aw}-q_b)^2/2,
	\quad \Gamma<2\omega_0, \quad n=1,2,3,...
	\nonumber
\end{eqnarray}

\noindent The flux barely changes near $t_n$ whereas it rises
sharply beyond this range provided the corresponding $n$ is not
too large \cite{infinity} (Fig.2(b)). In the underdamped case, the
\lq\lq length" of each step, $t_{n+1}-t_{n}$, is half a period of
eigenoscillation and the \lq\lq height" of the first steps is
large: $S(t_{n})- S(t_{n+1})$ $\approx \Delta U_p
\omega_0/(\pi\Gamma n(n+1))\stackrel{\Gamma\rightarrow
	0}{\longrightarrow} \infty$. As $\Gamma$ grows, the length of a
step increases while the height decreases and, at
$\Gamma=2\omega_0$, the steps vanish.

The instants $t_n$ mark intervals corresponding to different topologies
of the MPEP: for $t\leq t_1$, $[q(\tau)]$ is monotonic while, for
$t_n<t\leq t_{n+1}$ ($n=1,2,3,...$), $[q(\tau)]$ possesses $n$ turning
points. As $t$ changes, the MPEP varies {\it continuously} for any $t$,
including $t=t_n$. The exit velocity is non-zero unless $t=t_n$
(Fig.2(a)).

Apart from a quantitative description of the case when the wall is
close to the bottom of the well, the parabolic approximation
provides qualitative estimates of the time and energy scales of
the steps in the general case. However, some features of the steps
$S_{\rm min}(t)$ and of the associated evolution of the MPEP
change qualitatively as the absorbing wall moves towards the
saddle.

Let us move the absorbing wall $q_{aw}$ to a distinctly non-parabolic
region of $U(q)$, but still not too close to the saddle
($<q_{aw}^{(c)}$). One can reduce the 4th-order differential equation
(15) to a 2nd-order equation for $q$ plus a 1st-order one for the
auxiliary variable $\Gamma^{\prime}$ \cite{soskin}:
\begin{eqnarray}
	&& \ddot{q}+\Gamma^{\prime} \dot{q} +dU/dq=0
	\\
	&& [\dot{\Gamma}^{\prime} + (\Gamma^2-(\Gamma^{\prime})^2)/2]\dot{q}^2=2\Gamma
	\tilde{E},
	\nonumber
\end{eqnarray}

\noindent where
\begin {equation}
\tilde{E}\equiv -\frac{\partial S}{\partial t} =-L + \left(
\frac{\partial L}{\partial \dot{q}} -\frac{d}{dt}\left(\frac{\partial
	L}{\partial
	\ddot{q}}\right)\right)\dot{q}+\frac{\partial L}{\partial \ddot{q}}\ddot{q}
\end{equation}

\noindent is conserved along the MPEP \cite{elsgolc,soskin},
analogously to energy in mechanics \cite{landau}. Given that the
initial state is at the bottom, it can be shown that
$\tilde{E}\geq 0$ on the MPEP. Allowing for the fact that
$\partial S/
\partial \dot{q}(t)=0$ on the MPEP,

\begin {equation}
\frac{dS_{\rm min}}{dt}
=-\tilde{E}|_{\rm MPEP} \leq 0.
\end{equation}

The system (20), in addition to providing an algorithm \cite{footf}
that is faster in some ranges of parameters than solving Eq.\ (15), has
a remarkable feature: if $\tilde{E}=0$, the equation for
$\Gamma^{\prime}$ can be integrated explicitly \cite {soskin}. So, the
4th-order equation (15) reduces to a closed 2nd-order equation
\cite{footg}. Allowing for $\dot{q}_i=0$, the equation for the
time-reversed trajectory $[\tilde{q}(\tau)]\equiv [q(t-\tau)]$ becomes

\begin{eqnarray}
	&&
	\frac{d^2\tilde{q}}{d\tau^2}+\Gamma\frac{1+A{\rm e}^{\Gamma \tau}} {1-A{\rm e}^{\Gamma
			\tau}}\frac{d\tilde{q}}{d\tau} +\frac{dU(\tilde{q})}{d\tilde{q}}=0,
	\quad \quad A={\rm e}^{-\Gamma t}
	\nonumber
	\\
	&& \tilde{q}(0)=q_{aw}.
\end{eqnarray}

\noindent For the sake of convenience, we have also presented in (23)
the initial $\tilde{q}$ which follows from the third of conditions
(13). The derivative $ d\tilde{q}(\tau=0)/d\tau $ must be chosen such
that the condition (16) is satisfied: comparing Eq.(23) at $\tau=0$
with Eq.(16), we come to the important conclusion that

\begin {equation}
d\tilde{q}(\tau=0)/d\tau=0,
\end{equation}

\noindent i.e.\ the MPEP has a zero exit velocity if $dS_{\rm
min}/dt=0$.

One can show (cf.\ \cite{soskin}) that the number of possible
finite values of $t$ in eq.(23), such that $\tilde{q}(t)=q_b$,
equals the number $N$ of turning points in the noise-free
($t=\infty$) trajectory. Labelling such times $t$ as $t_n\equiv
t_n(q_{aw})$ ($n=1,2,...,N$), one may relate $n$ to the number
$n_{tp}$ of turning points in the trajectory (23)-(24):
$n=n_{tp}+1$. $t_n$ increases with $n$ and, if $N=\infty$, the
trajectory (23)-(24) for $t=t_n$ with $n\rightarrow \infty$
coincides with the noise-free trajectory. If

\begin {equation}
\Gamma<2\omega_0,
\end{equation}

\noindent then $N=\infty$ \cite{landau,soskin} while, if
$\Gamma\ge 2\omega_0$, then typically $N=0$. In rare cases, there
is a finite $N\ne 0$ at $\Gamma \geq 2\omega_0$ \cite {soskin}.

Thus, if $\Gamma<2\omega_0$, then $S_{\rm min}$ decreases with $t$
monotonically, possessing an infinite number of inflection points
$t_n$ with $dS_{\rm min}(t_n)/dt_n=0$ (Fig.3(a)). They divide the
time axis into intervals where the MPEP has different numbers of
turning points: as $t$ increases, the transformation of the MPEP
with $n-1$ turning points, into one with $n$ points, occurs {\it
continuously} at $t=t_n$.

At $\tau=0$, Eq.(23) coincides with the conventional relaxational
equation with a {\it finite} friction parameter, $\Gamma {\rm
	cth}(\Gamma t/2)$. Hence, the closer $q_{aw}$ is to the saddle,
the slower the motion near the wall. Thus, $t_n\rightarrow \infty$
if $q_{aw}\rightarrow q_s$. On the contrary, the time of motion
along MPEPs which get to the wall with non-zero velocity (they
relate to sections $S_{\rm min}(t)$ with non-zero $dS_{\rm
	min}/dt$) is less sensitive to the distance $q_s-q_{aw}$ and
remains finite even if $q_{aw}= q_s $. Consequently, as $q_{aw}$
grows, the onset of the {\it fold} at $t\approx t_1$ (according to
numerical calculations) occurs at the critical value
$q_{aw}^{(c)}$: $dS_{\rm min}/dt$ is discontinuous at the fold
(Fig.3(b)). At $q_{aw}>q_{aw}^{(c)}$, there are intervals of $t$
during which the system (13),(15)-(17) possesses more than one
solution \cite{footh}. It is because, the closer $q_{aw}$ is to
$q_s$, the larger is the number of such intervals and the maximal
possible number of coexisting solutions). This result provides a
{\it multi-branch} structure for $S(t)$ satisfying (13),(15)-(17)
(Fig.3(c)). In order to find the activation energy at a given $t$
one should choose from the solutions of (13),(15)-(17) the minimal
one. There are switches between different branches at certain
critical times. These can be compared to switching processes, as
other parameters vary, in certain escape problems
\cite{maierstein,soskin};
see also Sec. III.B below). The
switches result in {\it jump-wise} changes of the MPEP while the
activation energy still remains continuous (Fig.3(c)). At the same
time, the switch results in a discontinuity $dS_{\rm min}/dt$: its
values on different sides of the fold differ drastically, so that
$S_{\rm min}(t)$ and $J(t)$ are still distinctly step-wise
(Figs.3(c)).

We have tested some of the above predictions using computer
simulations. $S_{\rm min}(t)$ is derived via optimal fitting of $J(t)$
obtained at different $T$. Figs.\ 2(b) and 3 show reasonable agreement
between $S_{\rm min}(t)$ from the theory and from the simulations. The
growth of the flux is clearly step-wise (see insets) in both cases.

\begin{figure}
	\centering
	{\leavevmode\epsfxsize=3.2in\epsfysize=2in\epsfbox{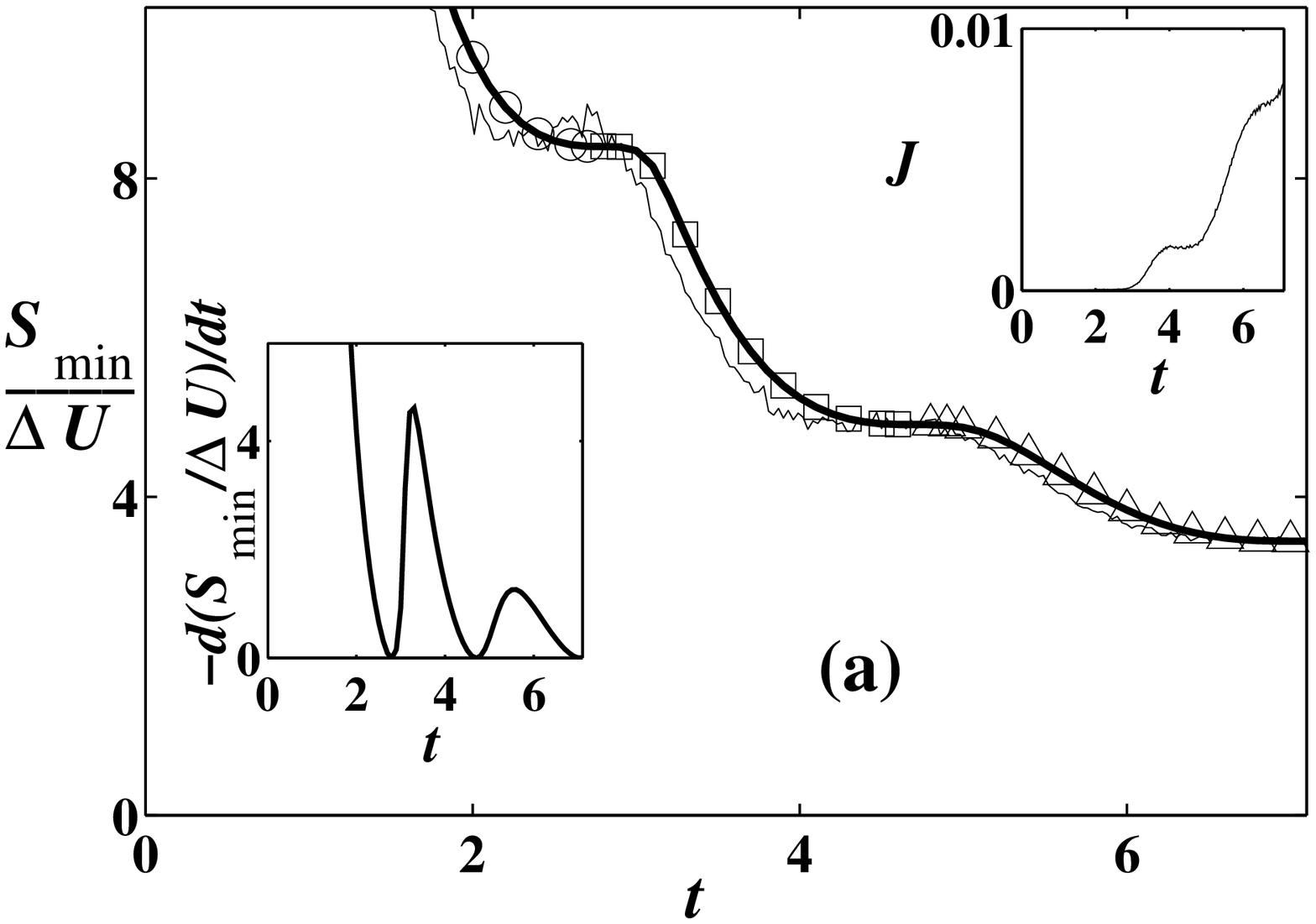}}
	{\leavevmode\epsfxsize=3.2in\epsfysize=2in\epsfbox{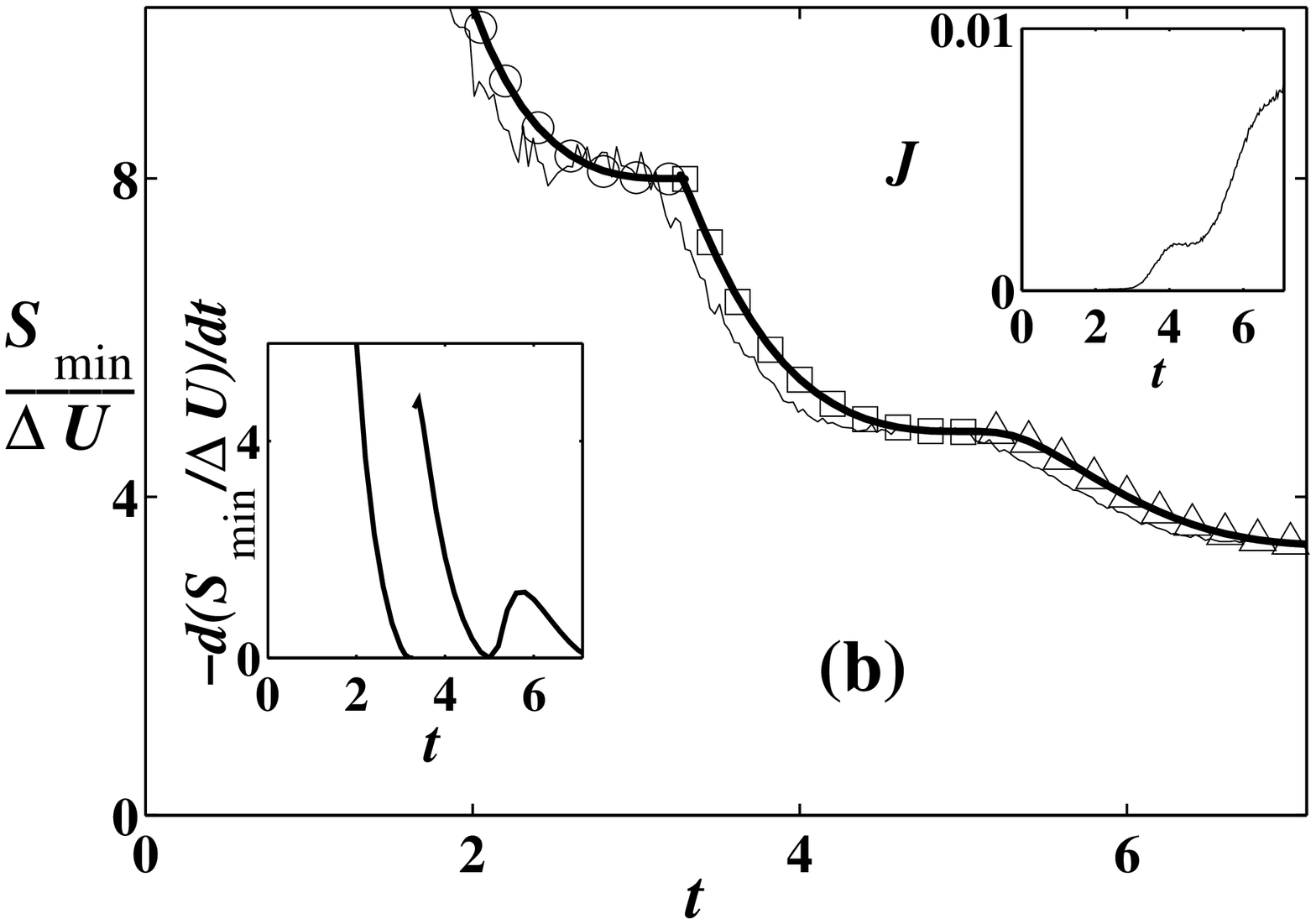}}
	{\leavevmode\epsfxsize=3.2in\epsfysize=2in\epsfbox{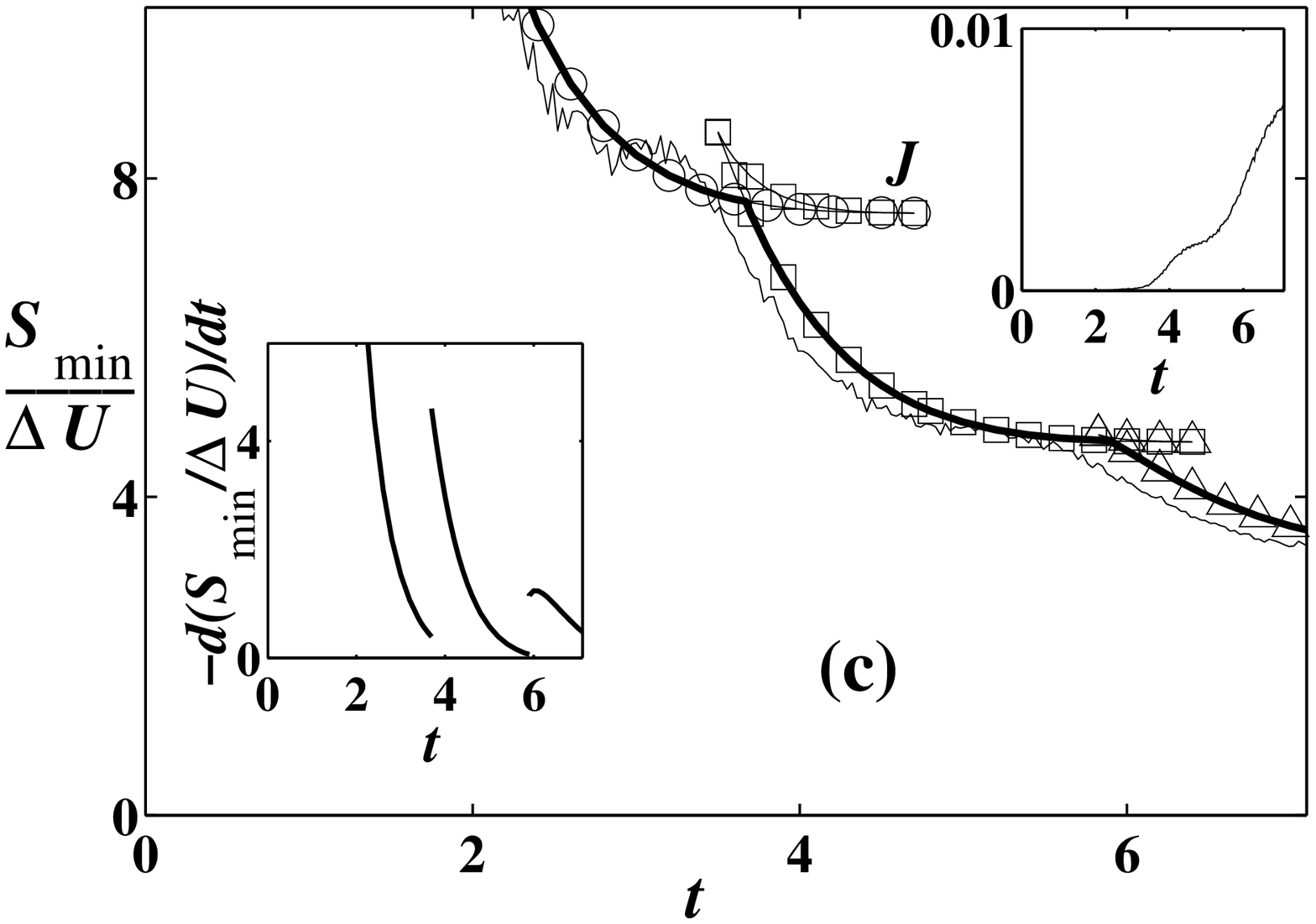}}
	\caption{The case of the bottom initial state. The evolution of $S_{\rm
			min}(t)$ normalized by $\Delta U\equiv U(q_{aw})- U(q_b)$ (thick and
		jagged lines for the theory and simulations respectively) as $q_{aw}$
		increases: (a) $q_{aw}=q_{aw}^{(2)}\equiv 0$, (b) $q_{aw}=0.371\approx
		q_{aw}^{(c)}$, (c) $q_{aw}=q_{aw}^{(3)}\equiv 0.9$. $\Gamma=0.05$.
		Branches of $S(t)$ corresponding to 0, 1 or 2 turning points in the
		escape path are shown by thin lines marked by circles, squares or
		triangles respectively: in (a) and (b), only one branch exists at each
		$t$ while, in (c), a few branches coexist in some ranges of $t$
		(activation energy $S_{\rm min}(t)$ coincides with the lowest $S(t)$).
		Left and right insets show respectively $-d(S_{\rm min}(t)/ \Delta U
		)/dt$ (theory) and $J(t)$ measured at $T=\Delta U$.}
\end{figure}

\subsubsection{2. Non-bottom initial state with a given coordinate and
	velocity}

If the initial state with a given coordinate and velocity,
$\{q_i,\dot{q}_i\}$, is shifted from the bottom of the well
$\{q_b,0\}$ then $ S_{\rm min}(t)$ changes: cf.\ Fig.4. Typically,
$ S_{\rm min}(t)$ becomes non-monotonic: cf. Fig. 4(b) (only if $\dot{q}_i=0$
it becomes step-wise i.e. monotonic: cf. Fig. 4(a)). Moreover, as is evident
in Fig.4, even a tiny shift of the energy from the bottom results
in quite a significant distortion of $ S_{\rm min}(t)$: the shift
of energy in Fig. 4(a) and Fig. 4(b) is equal to $\Delta U_p/100$
and $\Delta U_p/200$ respectively. Such strong sensitivity to the
initial state can be explained by the singularity in the effective
time-dependent damping parameter in equation (23), which describes
the MPEP; so, the shift in the activation energy depends
non-analytically on the shift of the energy of the initial state.

\begin{figure}
	\centering
	{\leavevmode\epsfxsize=3.2in\epsfbox{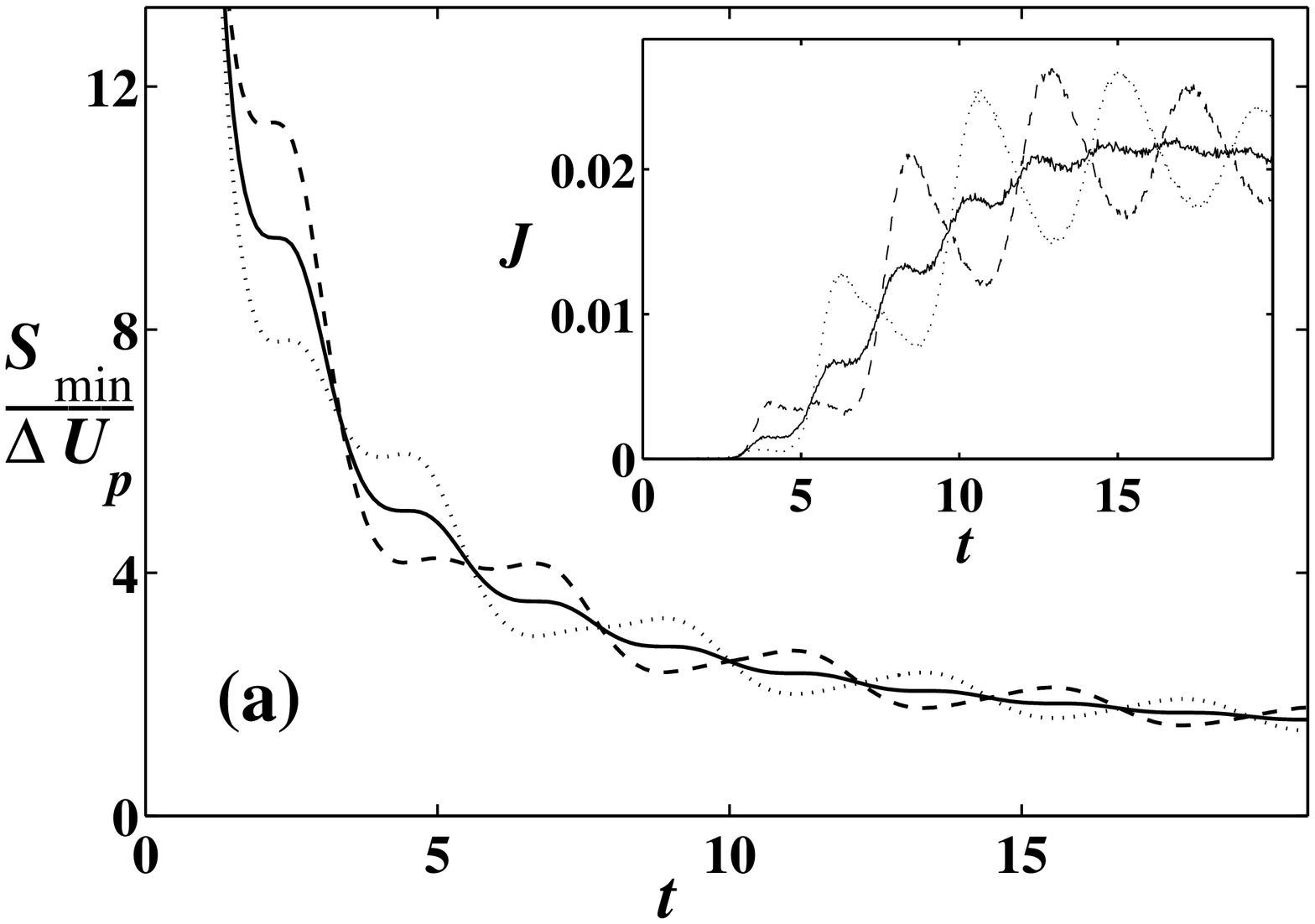}}
	{\leavevmode\epsfxsize=3.2in\epsfbox{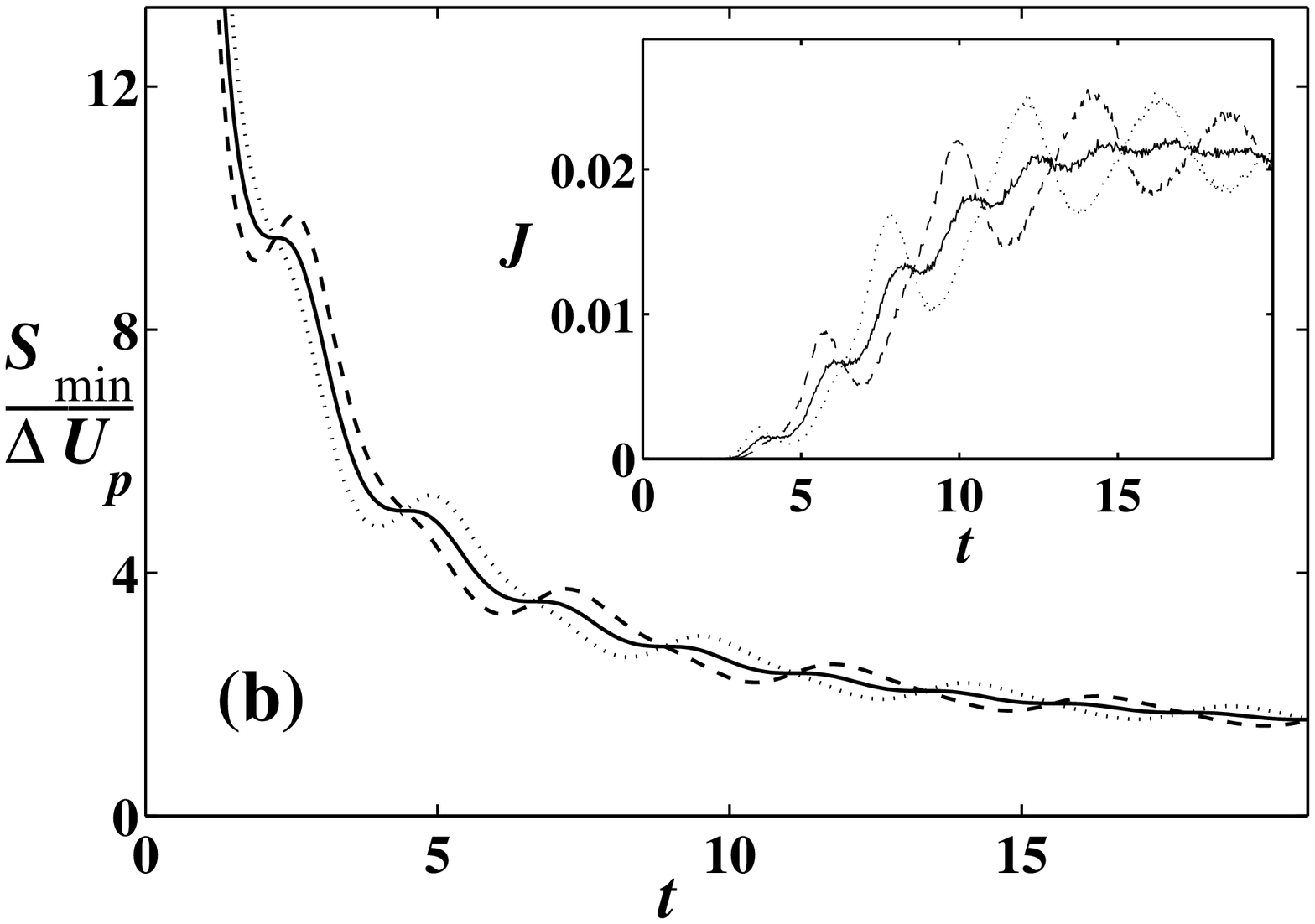}}
	\caption {Comparison between $S_{\rm min}(t)$ for the bottom initial
		state (solid line) and for two other initial states with given
		coordinate and velocity close to those in the bottom, with all other
		parameters the same as in Fig. 2: (a) $\dot{q}_0=0$ while
		$q_0=q_b-0.01$ (dotted line) or $q_0=q_b+0.01$ (dashed line); and (b)
		$q_0=q_b$ while $\dot{q}_0=-0.01$ (dotted line) or $\dot{q}_0=0.01$
		(dashed line).}
\end{figure}

\subsubsection{3. Thermalized initial state}

A non-bottom initial state with a given coordinate and velocity might
seem a rather artificial situation but, at the same time, there is
always some non-zero initial temperature $T_0$ so that various
non-bottom states are necessarily involved. The strong sensitivity of
the flux $ J_{q_i,\dot{q}_i} (t)$ to the shift of $\{q_i,\dot{q}_i\}$
from the bottom, appears to cast doubt on the generality of the
stepwise growth $J$ in real situations. However, a rigorous analysis
(see below) shows that the flux at short time scales still grows in a
stepwise manner for any temperature $T_0<T$. Moreover, if $T_0/T\ll
\Gamma/\omega_0$, then the step-wise structure for flux growth is
similar to that obtained using the bottom as the initial state.

So, let the distribution of initial coordinates and velocities be
quasi-stationary for some low temperature $T_0$:

\begin{eqnarray}
	&& W_{in}(q_{0},\dot{q}_0)\approx
	 W_{qs}(q_{0},\dot{q}_0)
	 \equiv
	\nonumber
	\\
	&& \equiv \left\{
	^{\exp(-E_0/T_0)/Z \quad
		{\rm for}\quad E_0<U(q_{aw}),}
	_{0
		\quad\quad \quad\quad \quad\quad {\rm for}\quad E_0>U(q_{aw}),}
	\right.
	\\
	&& E_0\equiv
	\dot{q}_0^2/2+U(q_0),
	\nonumber
	\\
	&& Z=\int\int_{ E_0<U(q_{aw})}dq_0d\dot{q}_0 \; \exp(-E_0/T_0).
	\nonumber
\end{eqnarray}

\noindent We assume that the probability for the system to leave the
well before the relevant \lq\lq initial'' instant $t=0$ is negligible.

If at the \lq\lq initial'' instant $t=0$ the additional noise source is
switched on, so that the effective temperature becomes $T>T_0$
\cite{footnontherm}, the evolution of the flux (6) with the
initial distribution (26) becomes relevant. Given the activation-like
structure of $ J_{q_0,\dot{q}_0}(t)$ (eqs. (10)-(13)), the flux with
the thermalized initial state can be presented in the form

\begin{equation}
	J(t)\equiv J_{T_0}(t)=\tilde {P} {\rm e}^{-\frac{\tilde {S}_{\rm min} (t)}{T}}
\end{equation}

\noindent where $\tilde {P} $ is some prefactor and $\tilde {S}_{\rm
	min} $ is the generalized activation energy:

\begin{equation}
	\tilde {S}_{\rm min}
	\equiv \tilde {S}_{\rm min}
	\left(\frac{T_0}{T},t\right)
	= {\rm min}_{q_0,\dot{q}_0}
	\left\{
	S_{\rm min}(q_0,\dot{q}_0,t) +
	\frac{T}{T_0}E_0
	\right\},
\end{equation}

\noindent where $ S_{\rm min}(q_0,\dot{q}_0,t) $ is given by (11)-(13)
and $E_0$ is defined in (26).

There is no room here to provide details but it can be shown that, for
any $T_0<T$, the function $\tilde {S}_{\rm min} (\frac{T_0}{T},t)$ is
stepwise in $t$. Analogously to the case of the bottom initial state,
$\tilde {S}_{\rm min} $ possesses inflection points with $d\tilde
{S}_{\rm min}/dt=0$, provided the wall is not too close to the saddle,
and the corresponding MPEPs are described by an equation similar to
(23) but with the constant $A$ related to $t$ as

\begin{equation}
	A={\rm e}^{-\Gamma t}\left(1-\frac{T_0}{T}\right).
\end{equation}

\noindent The relevant instants $t$ are determined using the condition
$\dot{\tilde{q}}(t)=0$, rather than the condition $\tilde{q}(t)=q_b$
which is relevant to the bottom initial state.

\begin{figure}
	\centering
	{\leavevmode\epsfxsize=3.2in\epsfbox{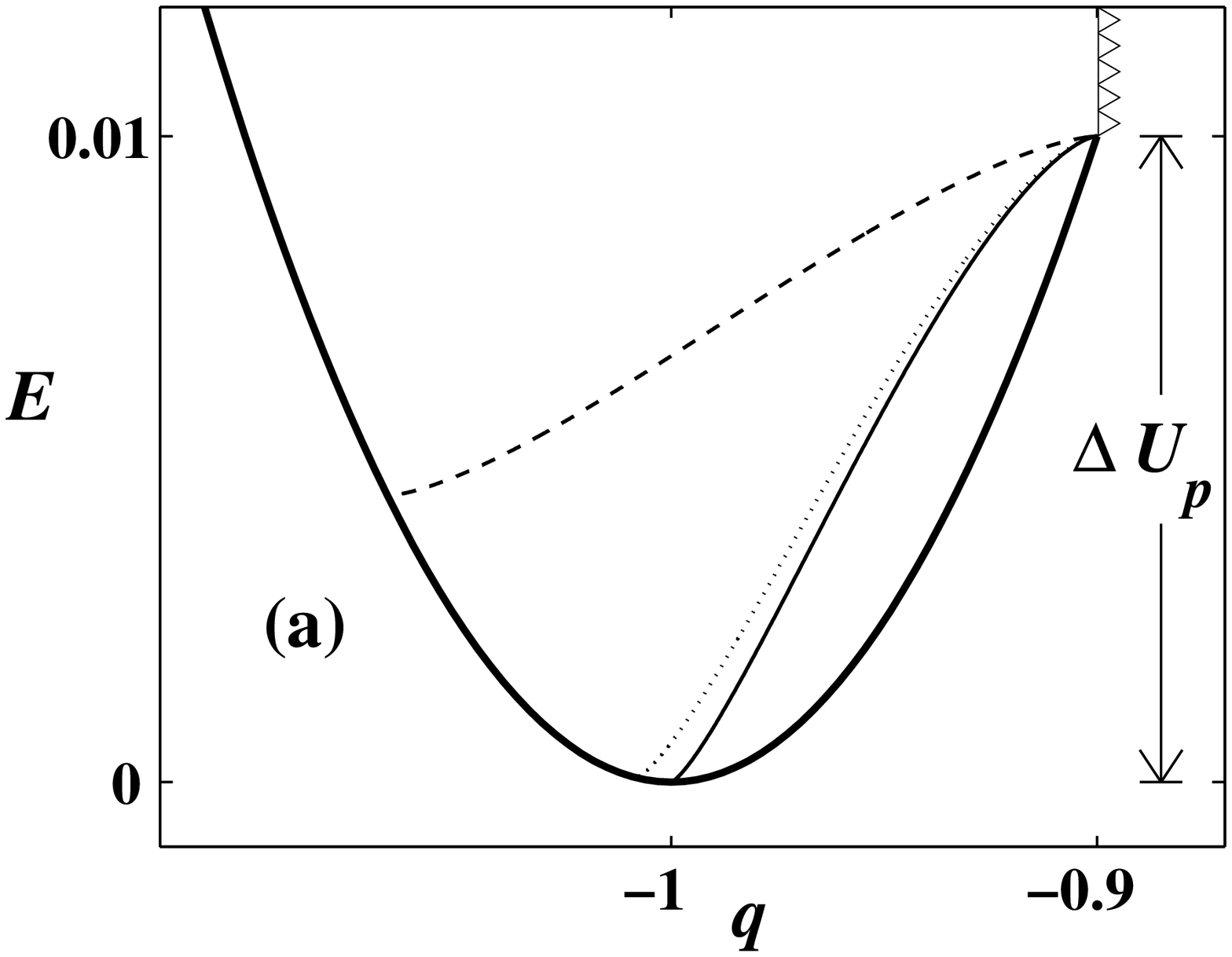}}
	{\leavevmode\epsfxsize=3.2in\epsfbox{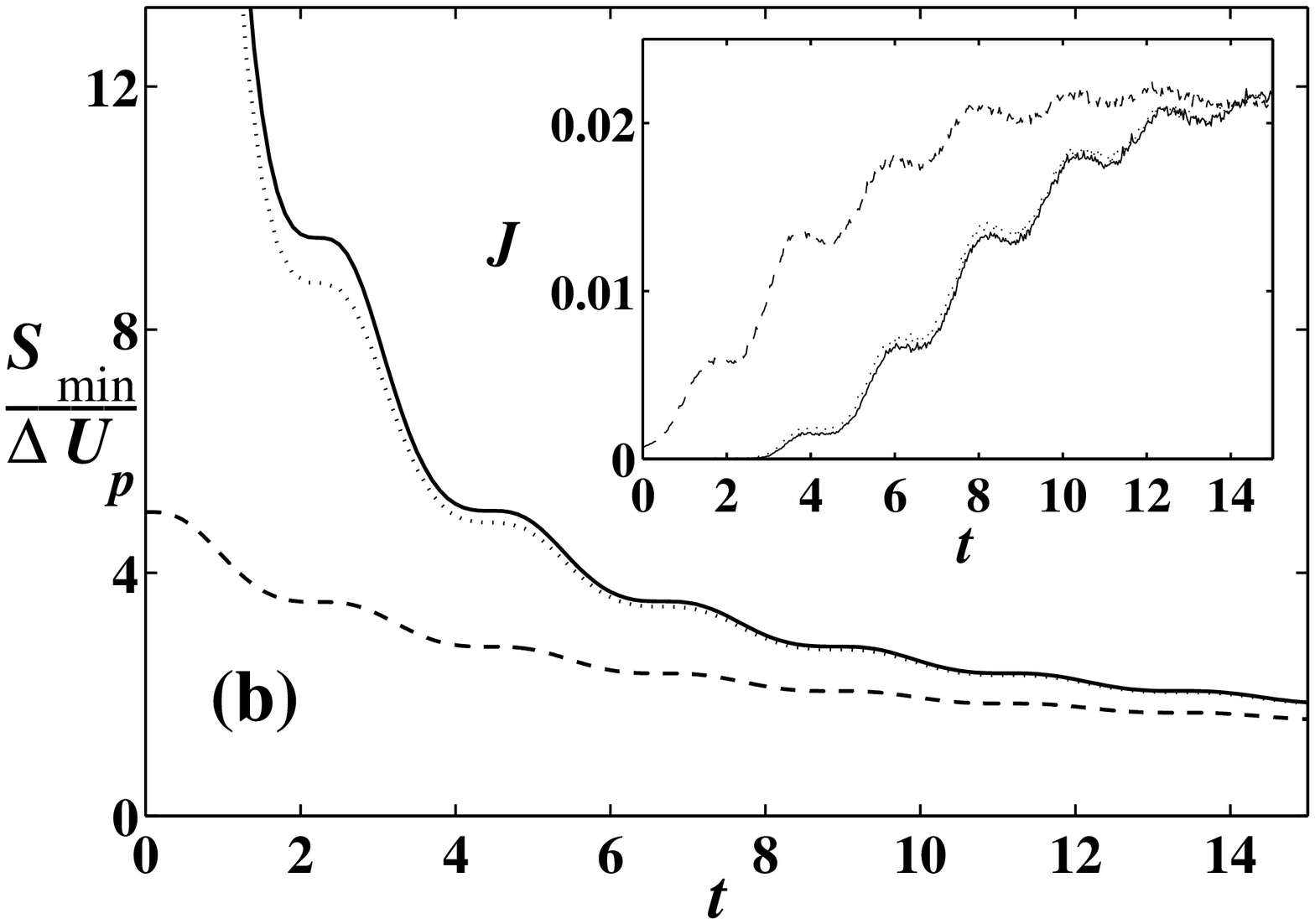}}
	\caption {The case of the thermalized initial state. (a) MPEPs for $t=2.222$, for three characteristic values of
		$T_0/T$, with all other parameters the same as in Fig. 2: $T_0/T=0$
		(solid line), 0.01 (dotted line), 0.2 (dashed line); (b) $\tilde
		{S}_{\rm min}(t)$ for $T_0/T=0$ (solid line), 0.01 (dotted line), 0.2
		(dashed line).}
	
\end{figure}

It can be shown that $\tilde {S}_{\rm min}
(\frac{T_0}{T},t \sim\omega_0^{-1})$ is close to
$ S_{\rm min}(q_i=q_b,\dot{q}_i=0,t\sim\omega_0^{-1}) $ provided

\begin{equation}
	\frac{T_0}{T}\ll\frac{\Gamma}{\omega_0}.
\end{equation}

\noindent Otherwise $\tilde {S}_{\rm min}(t\sim\omega_0^{-1})$ is
significantly lower and the steps are smeared (Fig. 5(b)).

The competition between the two small parameters, $T_0/T$ and
$\Gamma/\omega_0$, is readily interpreted physically. On one hand, the
escape flux (on $t\sim\omega_0^{-1} $) from the bottom is $\propto
\exp(-a\Delta U/(T\Gamma/\omega_0))$ where $a\equiv a(t)\sim 1$. On the
other hand, if the system starts its motion from an energy $E_0$ close
to the barrier level, the probability of escape for time $t\sim
\omega_0^{(-1)}$ will be $\sim 1$, but then the probability to have
such starting energy is $\propto \exp(-\Delta U/T_0)$. It is the
competition between these two exponentially weak processes which leads
to the relation (30). Fig. 5(a) shows that, for $T_0/T=0.01\ll
\Gamma/\omega_0\approx 0.035$, the MPEP starts close to the bottom
while, for $T_0/T=0.2\gg \Gamma/\omega_0$, the starting energy is $\sim
\Delta U$.

\subsection{B. Multi-well metastable potential}

\begin{figure}
	\centering
	{\leavevmode\epsfxsize=3.3in\epsfysize=2.6in\epsfbox{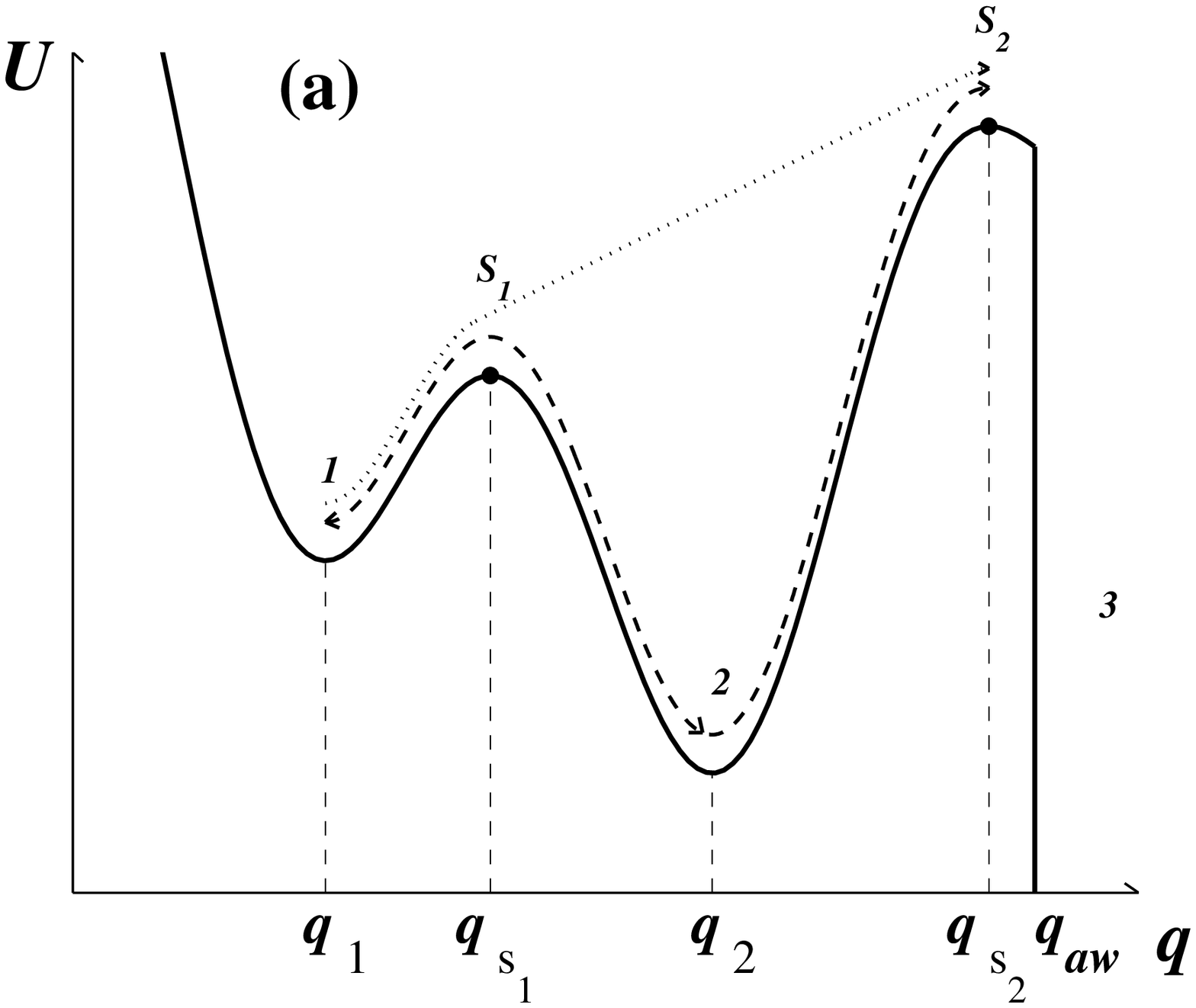}}
	{\leavevmode\epsfxsize=3.3in\epsfysize=2.6in\epsfbox{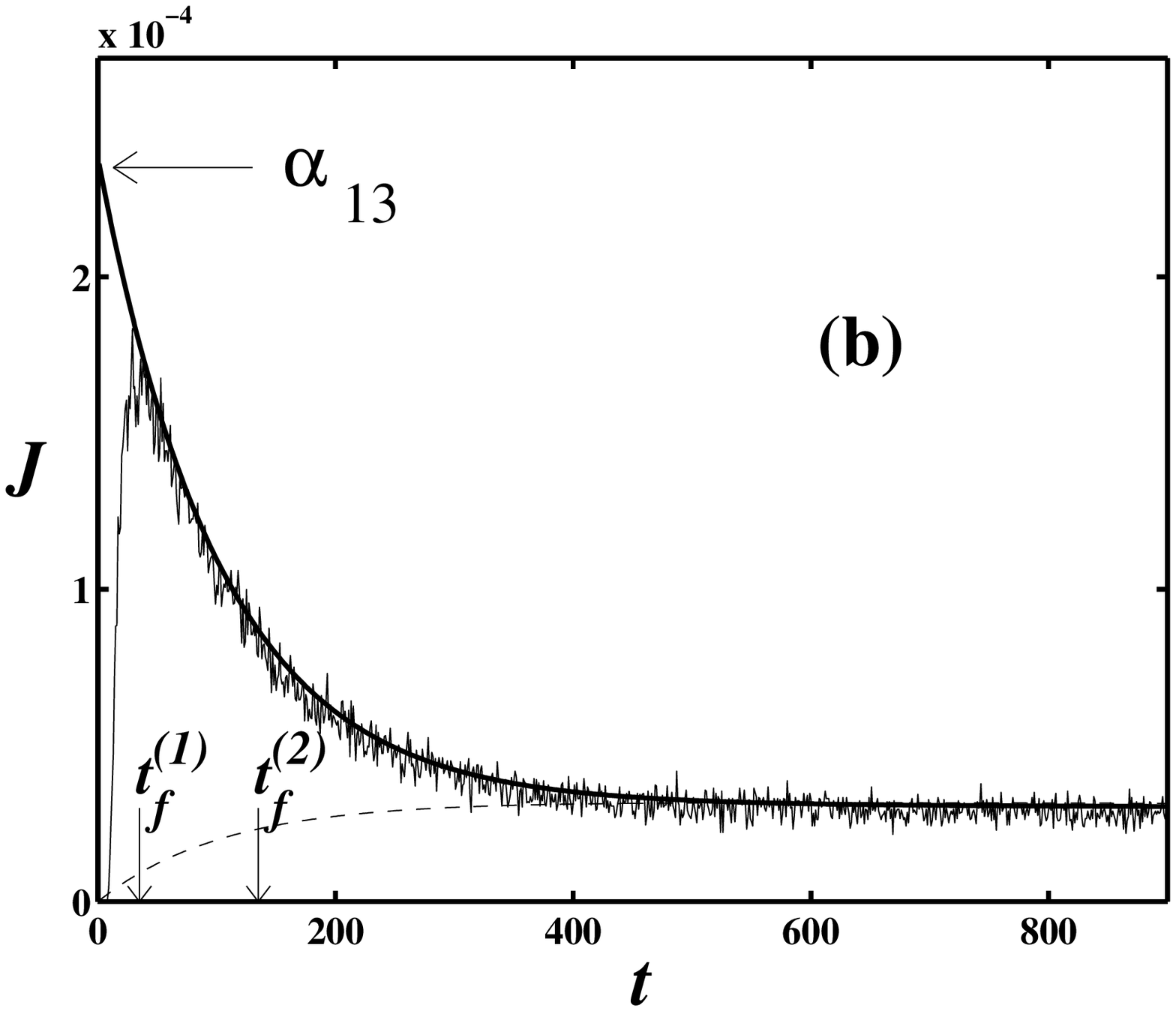}}
	\caption{(a) the potential (4) and a sketch of direct (dotted line) and
		indirect (dashed line) escape paths $1\rightarrow {\rm s}_2$; thin
		dashed lines indicate positions of the local minima ($q_1, q_2$) and
		maxima ($q_{\rm s_1}, q_{\rm s_2}$); (b) simulations of the dependence of
		the escape flux on time $J(t)$ (thin line) for the model (1),(4) at
		$\Gamma=0.15$, $T=0.4$. The thick full and dashed lines show the
		approximation of $J(t)$ by eq.(31) in which $\alpha_{12}$,
		$\alpha_{21}$, $\alpha_{qs}$ are calculated by the Kramers-Melnikov
		formula
		\cite{Melnikov:91}.
		For the thick full line,
		$\alpha_{13,23}=\alpha_{qs}(1+\{\Omega_1\Omega_2^{-1}\exp
		[(U_1-U_2)/T]\}^{\pm 1})/(1+\{m$ $\exp [k S_{\rm min} ({\rm
			s}_2\rightarrow {\rm s}_1)/T]\}^{\pm 1})$ where $\Omega_{1,2}$ are the
		frequencies of eigenoscillation in the bottom of wells 1,2
		respectively, $k$ is equal to 1,-1 for the ranges $\Gamma$ providing
		${\rm s}_2\stackrel{nf}{\rightarrow}2,1$ respectively, $S_{\rm min}
		({\rm s}_2\rightarrow {\rm s}_1)$ is calculated from the theory
		\cite{soskin}
		and
		$m$ is the only adjustable parameter ($m\approx 1.1$ for these
		parameters); for the dashed line, $\alpha_{13}=0$ and
		$\alpha_{23}=\alpha_{qs}(1+\alpha_{21}/\alpha_{12})$.}
\end{figure}

As an example of the multi-well case, we consider the potential (4),
which describes the simplest SQUID \cite{likharev}. We place an
absorbing wall \cite{footb} at $q_{aw}=4.5$ (Fig.6(a)) while the
initial state of the system (1),(4) may be any state within well-1; in
simulations, we put it at the bottom of well-1, for the sake of
simplicity. We emphasize also that the type of the boundary is not
important either, e.g.\ our results are equally valid for the
transition rates between non-adjacent wells in the stable potential
with more than two wells \cite{soskin}.

Unlike the single-well case, where the formation time of
quasi-equilibrium is of the order of $ t_f^{(s)}$ (7), its formation in
the multi-well case proceeds via two distinct stages: first,
quasi-equilibrium is formed within the {\it initial well} which takes
$t_f^{(1)}\sim t_f^{(s)}$: $J$ evolves at this stage quite similarly
\cite{additional} to the single-well case; secondly, quasi-equilibrium
{\it between wells} becomes established which takes exponentially
longer: $t_f^{(2)}\sim t_f^{(s)} \exp (\Delta U / T)\gg t_f^{(1)}$
where $\Delta U $ means a minimal internal barrier. During the latter
stage, and during the subsequent quasi-stationary one, the flux $J(t)$
can be described via a solution of kinetic equations for the well
populations, $W_1$ and $W_2$, using the concept of constant
inter-attractor \cite{footi} transition rates $\alpha_{ij}$
(cf.~\cite{freidlin84}):

\begin{eqnarray}
	&&J(t)\equiv W_1\alpha_{13}+W_2\alpha_{23}=
	\\
	&&\quad\quad\quad\quad\quad\quad
	\alpha_{13} {\rm e}^{-\frac {t} {t_f^{(2)}}} +  \alpha_{qs} \left({\rm e}^{-
		\frac {t} {t_{qs}}}-
	{\rm e}^{-\frac {t} {t_f^{(2)}}}\right),
	\nonumber \\
	&&
	t_f^{(2)}\approx \alpha_{12}^{-1}, \quad\quad
	t_{qs}\approx \alpha_{qs}^{-1}\approx
	\alpha_{12}/(\alpha_{12}\alpha_{23}+\alpha_{21}\alpha_{13}),
	\nonumber
	\\
	&&
	T\ll U_{\rm s_1}-U_1, \quad \quad t\gg t_f^{(1)}.
	\nonumber
\end{eqnarray}

\noindent The physical meaning of the two terms in (31) is easily
understood (cf.~Fig.~6). The first one corresponds to {\it direct}
escapes, i.e.~those that do not go via the bottom of well-2, and it
dominates until quasi-equilibrium becomes established. The second term,
corresponding to indirect escapes, i.e.\ those that involve one or more
intermediate transitions between wells 1 and 2 while the ultimate
transition to 3 may occur from either well. It dominates during the
ensuing quasi-stationary stage: it is the asymptotic part of this
latter flux, $\alpha_{qs}\exp(-t/t_{qs})$, that is called the
quasi-stationary flux.

Thus, in order to know the flux dynamics one needs to find the
inter-well transition rates $\alpha_{ij}$. The rates $\alpha_{12},
\alpha_{21}$ and the quasi-stationary rate $\alpha_{qs}$ can be
calculated from the Kramers-Melnikov formula \cite {Melnikov:91}. Thus, only
one of the four $\alpha_{ij}$ coefficients needs to be found
independently. We choose $\alpha_{13}$ as the independent coefficient.

The theoretical problem of finding $\alpha_{13}$ is inherently
difficult. Melnikov pointed out \cite {Melnikov:91} that his method is valid
in the multi-well case only if the barriers levels are equal or at
least close to each other (cf.\ e.g.\ \cite{Melnikov:91,new}), a requirement
that is often not satisfied. So, the method of {\it optimal fluctuation}
(cf.\ the previous section) was suggested \cite {soskin}, seeking the
escape rate in the form

\begin{equation}
	\alpha_{13}=P{\rm e}^{-\frac{S_{\rm min}}{T}},
\end{equation}

\noindent where the action $ S_{\rm min}$ does not depend on $T$ and
the dependence of the prefactor $P$ on $T$ is relatively weak.

\begin{figure}
	\centering
	{\leavevmode\epsfxsize=3.3in\epsfbox{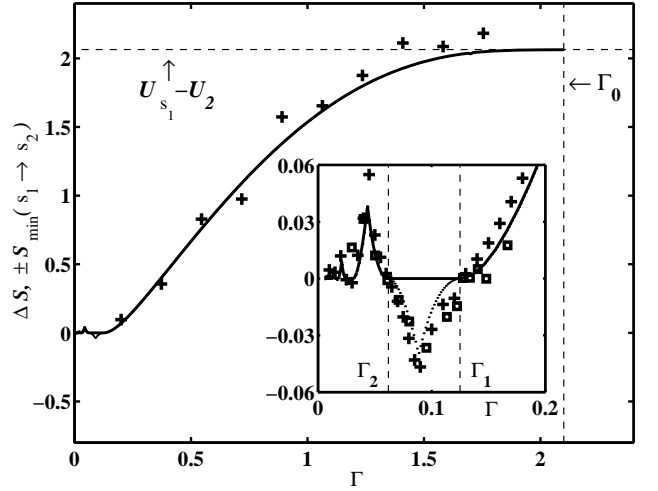}}
	\caption{Theoretical and experimental data on direct
		escapes/transitions in the metastable potential (4) (Fig.4(a)). The
		calculated excess of action
		over a difference of energies, $\Delta S(1\rightarrow {\rm s}_2)$ (34),
		is shown by the full line. It is related to the escape rate $\alpha_{13}$.
		The calculated $\pm S_{\rm min}({\rm
			s}_2\rightarrow {\rm s}_1)$, related to $R$ (37) by Eq.(38), is shown
		by the dotted line. It overlaps the full line in the half-plane of
		positive ordinates. The corresponding quantity (39) based on data
		obtained by electronic and computer simulations is shown by squares and
		crosses respectively. Values of $\Gamma_{n\ge 1}$ correspond to noise-free
		saddle-connections with $n-1$ turning points. At
		$\Gamma=\Gamma_0=2\Omega_2\approx 2.1$, the turning points in the
		noise-free trajectories ${\rm s}_2\stackrel{nf}{\rightarrow}2$ and
		${\rm s}_1\stackrel{nf}{\rightarrow}2$ disappear. The inset shows the
		low $\Gamma$ range enlarged.}
\end{figure}

\begin{figure}
	\centering
	{\leavevmode\epsfxsize=3.3in\epsfysize=2.2in\epsfbox{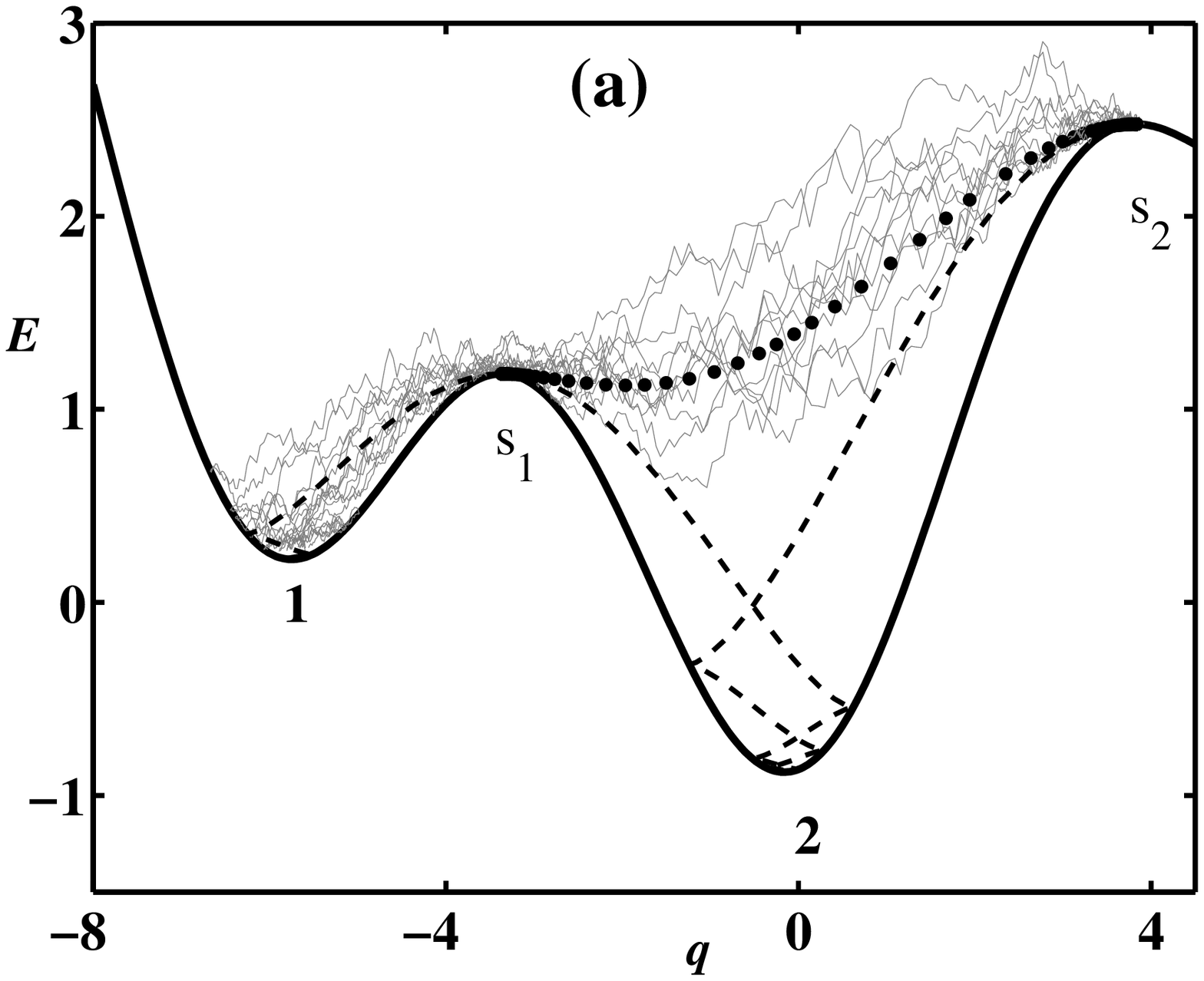}}
	{\leavevmode\epsfxsize=3.3in\epsfysize=2.2in\epsfbox{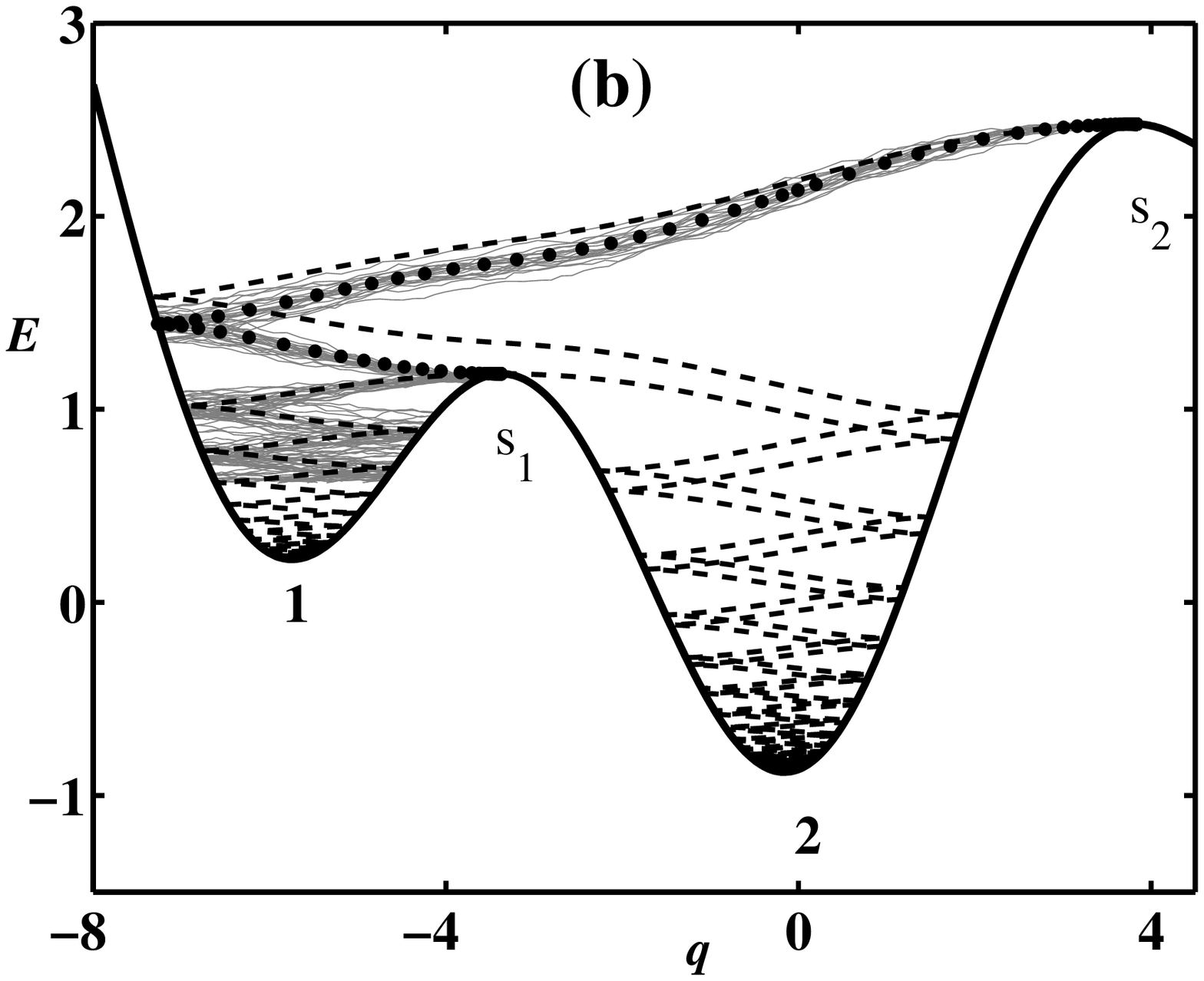}}
	{\leavevmode\epsfxsize=3.3in\epsfysize=2.2in\epsfbox{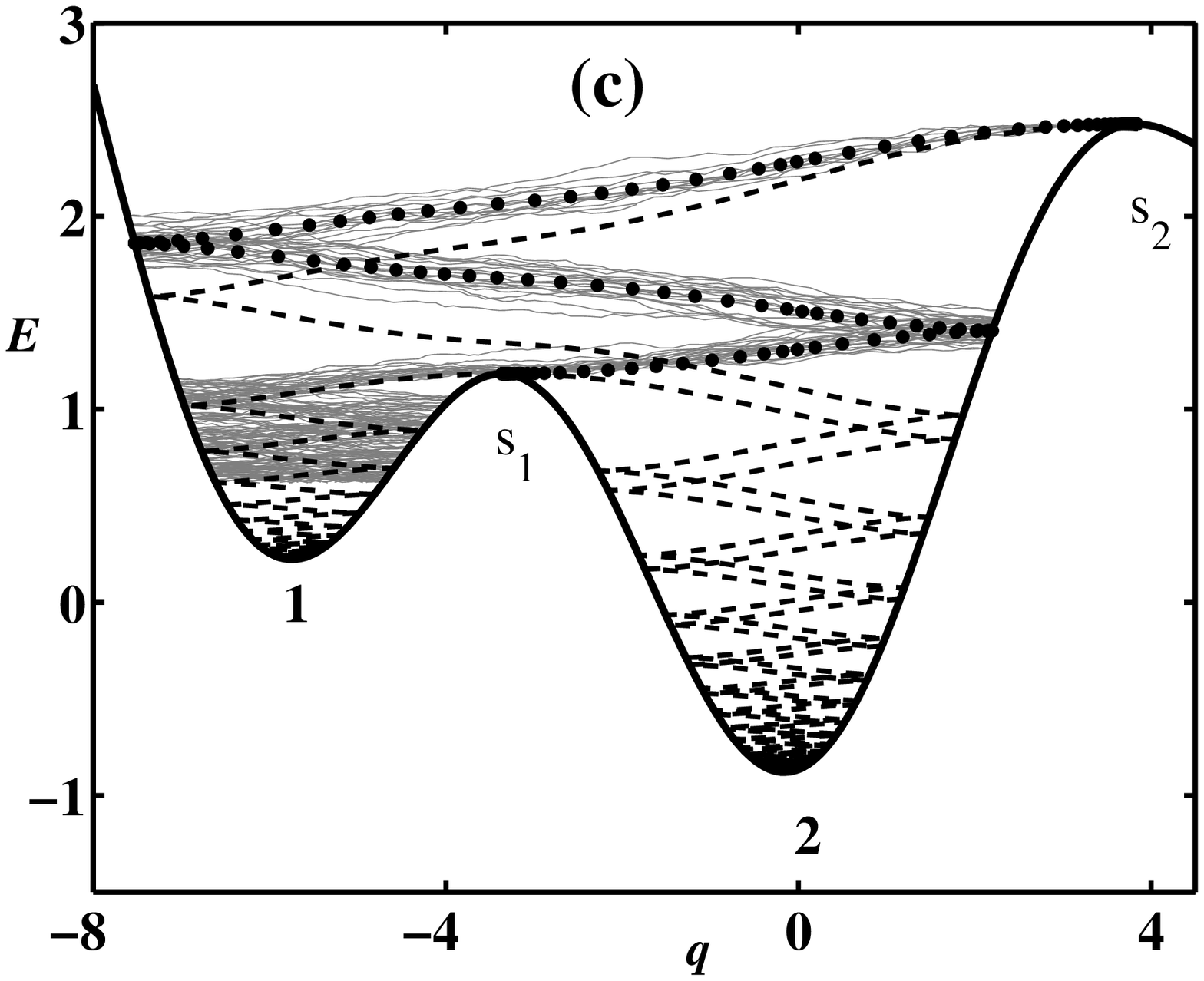}}
	\caption{Simulated direct transition paths ${\rm s}_2\rightarrow 1$
		(thin full lines) in the energy-coordinate plane $E-q$ (where
		$E=\dot{q}^2/2+U(q)$) corresponding to (1),(4) at different $\Gamma$:
		(a) $0.5$, (b) $0.05$, (c) $0.04$ ($T=0.05$ for (a) and $T=0.005$ for
		(b), (c)). The noise-free trajectories ${\rm
			s}_2\stackrel{nf}{\rightarrow}2$ and ${\rm
			s}_1\stackrel{nf}{\rightarrow}1,2$ are shown by dashed lines. The
		MPDTPs ${\rm s}_2\rightarrow {\rm s}_1$ are shown by thick dotted
		lines.}
\end{figure}

One can show that $ S_{\rm min}$ is the minimum of a certain functional
\cite{soskin}

\begin {eqnarray}
\label{activation}
& &
S_{\rm min} \equiv S_{\rm min}(1 \rightarrow {\rm s}_2)
= {\rm min}_{[q(t)],t_{tr}} (S), \\
& &
S\equiv S_{t_{tr}}[q(t)]= \frac {1}{4\Gamma}
\int_0^{t_{tr}} dt (\ddot {q} +\Gamma \dot {q} + dU/dq)^2,
\nonumber
\\
& &
q(0)=q_1,\quad \dot {q}(0)=0, \quad
q(t_{tr})=q_{{\rm s}_2},\quad \dot {q}(t_{tr})=0,
\nonumber
\end{eqnarray}

\noindent where the trajectory $[q(t)]$ does not pass through attractor
2. It can easily be shown that the $t_{tr}$ yielding $S_{\rm min}$ is equal
to $\infty$. The $[q(t)]$ yielding $S_{\rm min}$ is called
\cite{soskin} the {\it most probable direct transition path} (MPDTP).
The main features of $S_{\rm min}$ and the MPDTP are illustrated in
Figs.~7 and 8 for the system (1),(4); see \cite{soskin} for a rigorous
general treatment \cite{footj}.

Fig.\ 7 shows how the excess action

\begin{equation}
	\Delta S \equiv \Delta S (1\rightarrow {\rm s}_2) =S_{\rm min}(1\rightarrow {\rm
		s}_2) - (U_{{\rm s}_2}-U_1)
\end{equation}

\noindent varies with $\Gamma$ over the whole range of $\Gamma$,
from very strong damping to the ultra-underdamped case. One can resolve
three distinct regions.

The overdamped region can be defined as $\Gamma \ge \Gamma_0=2\Omega
_2$, where $\Omega _2$ is the frequency of eigenoscillation in the
bottom of well--2. Here, there is no MPDTP $1\rightarrow {\rm s}_2$ at
all, so that $\alpha_{13}=0$.

In the moderate-friction region, $[\Gamma_1, \Gamma_0]$, $\Delta
S(\Gamma)$ is monotonic and undergoes its largest variation: from $0$
to $U_{{\rm s}_1}-U_2$. The MPDTP (see Fig.~8(a)) is the time-reversed
trajectory ${\rm s}_2\stackrel{A=A_-}{\rightarrow} {\rm
	s}_1\stackrel{nf}{\rightarrow}1 $ in which the latter is just the
noise-free relaxation from ${\rm s}_1$ to 1, whereas the former is the
solution (cf. \cite{gt} and the previous section) of

\begin{eqnarray}
	\label{saddleconnection}
	&&
	\ddot {q}_d +\Gamma \frac {1+Ae^{ \Gamma t}}{1-Ae^{ \Gamma t}}\dot {q}_d +
	dU(q_d)/dq_d =0, \\
	&&
	q_d(0) =q_{{\rm s}_2}, \quad\quad \dot {q}_d(0)=0,  	
	\nonumber \\
	&&	q_d(t\rightarrow \infty) \rightarrow q_{{\rm s}_1}, \quad\quad \dot {q}_d(t\rightarrow \infty) \rightarrow 0.
	\nonumber
\end{eqnarray}

\noindent  Here $A=A_-$ is a negative constant providing for the minimal
$S$ among all values of $A$ for which $[q_d(t)]$ reaches ${\rm s}_1$.
Note that, in general, there may be an infinite set of $A$ providing
$[q_d(t)]$ connecting the saddles: the corresponding trajectories
differ by their number of turning points.

The underdamped region, $\Gamma \le \Gamma_1$, is divided by a number
of characteristic values of the friction $\Gamma_{n\ge 1}$. Each of
these $\Gamma_n$ provides for a {\it noise-free} saddle-connection
${\rm s}_2\stackrel{nf}{\rightarrow}{\rm s}_1 $, which possesses $n-1$
turning points. In this region, $\Delta S(\Gamma)$ undergoes
oscillations corresponding to an alternation between two situations. In
the first, $[\Gamma_{2m}, \Gamma_{2m-1}]$ ($m\ge 1$), a noise-free
trajectory ${\rm s}_2\stackrel{nf}{\rightarrow}1$ exists and the MPDTP
is just its time-reversal, with $\Delta S =0$. In the second situation,
$[\Gamma_{2m+1}, \Gamma_{2m}]$ ($m\ge 1$), the action varies
nonmonotonically with $\Gamma$, and has cusps. This is due to a
competition between the two paths which are the time-reversals
respectively of ${\rm s}_2\stackrel{A_-}{\rightarrow} {\rm
	s}_1\stackrel{nf}{\rightarrow}1 $ and ${\rm
	s}_2\stackrel{A_+}{\rightarrow} {\rm s}_1\stackrel{nf}{\rightarrow}1 $,
where ${\rm s}_2\stackrel{A_{\pm}}{\rightarrow} {\rm s}_1$ are given by
the solutions of (35) with $A_+\equiv A_+(\Gamma)>0$ and $A_-\equiv
A_-(\Gamma)<0$ respectively: see Fig.8(b) and Fig.8(c) respectively. As
$\Gamma$ varies, $S$ along one path becomes equal to $S$ along another,
at a certain $\Gamma$, leading to switching between the paths and to
the cusp in $\Delta S (\Gamma)$: there are corresponding
discontinuities in the non-equilibrium potential \cite{gt} and
fluctuational separatrix \cite{pl94}.

Thus, \cite{soskin} predicts an exponentially strong dependence of the
escape rate $\alpha_{13}$ on friction, including interesting features
such as oscillations and cusps \cite {others}, for $t\gg t_f^{(1)}$. To
establish whether these, and the properties of MPDTPs described above
occur in reality, we have undertaken analogue electronic and computer
simulations. A necessary condition is smallness of the temperature:
$T\ll \Delta S,(U_{{\rm s}_1}-U_1)$. However to obtain reasonable
statistics at such a small temperature would require an unrealistically
long time ($\propto \exp((U_{{\rm s}_2}-U_1+\Delta S)/T)$)
\cite{footk}. We have overcome this difficulty by exploiting the
property of detailed balance \cite{fpe}, which implies \cite{soskin}
that the MPDTP ${\rm s}_2\rightarrow 1$ is just the time-reversal of
the MPDTP $1\rightarrow {\rm s}_2$, with the corresponding actions
differing by $U_{{\rm s}_2}-U_1$ i.e.\ \begin{equation} \Delta S
	(1\rightarrow {\rm s}_2) = S_{\rm min}({\rm s}_2\rightarrow 1)= \{^{0
		\quad\quad\quad \quad\quad{\rm at}\quad {\rm
			s}_2\stackrel{nf}{\rightarrow}1,}_{S_{\rm min} ({\rm s}_2\rightarrow
		{\rm s}_1) \quad{\rm at}\quad {\rm s}_2\stackrel{nf}{\rightarrow}2,}
\end{equation}

\noindent so that information about the transition ${\rm
	s}_2\rightarrow 1$ is equivalent to that for $1\rightarrow {\rm s}_2$,
but the experimental time required is of course much smaller in the
former case ($\propto \exp(\Delta S/T)$) than in the latter.

Fig.8(a) demonstrates that, for $\Gamma \in [\Gamma_1, \Gamma_0]$, most
of the direct paths ${\rm s}_2\rightarrow 1$ do indeed concentrate near
${\rm s}_2\stackrel{A_-}{\rightarrow} {\rm
	s}_1\stackrel{nf}{\rightarrow}1$. Figures 8(b) and 8(c) demonstrate
switching of the MPDTP from ${\rm s}_2\stackrel{A_+}{\rightarrow} {\rm
	s}_1\stackrel{nf}{\rightarrow}1$ to ${\rm
	s}_2\stackrel{A_-}{\rightarrow} {\rm s}_1\stackrel{nf}{\rightarrow}1$
as $\Gamma$ decreases in the range $[\Gamma_3, \Gamma_2]$.

In order to study $S_{\rm min}$ we use the following technique. The
system is put at ${\rm s}_2$, and one then follows its stochastic
dynamics (1),(4) until either the bottom of one of the wells is
approached or the coordinate $q_{aw}$ is reached. After that, the
system is reset to ${\rm s}_2$ and the operation is repeated. Once
adequate statistics have been obtained, we calculate the ratio of
transitions to wells 1 and 2 respectively:

\begin{equation}
	R\equiv R(T) = \frac{N_{{\rm s}_2\rightarrow 1}}{ N_{{\rm s}_2\rightarrow 2}}.
\end{equation}

\noindent It is easy to see that $ R\propto \exp (\pm S_{\rm min} ({\rm
	s}_2\rightarrow {\rm s}_1)/T)$ (where $+,-$ correspond to ranges of
$\Gamma$ providing ${\rm s}_2\stackrel{nf}{\rightarrow}1,2$
respectively). So, $S_{\rm min} ({\rm s}_2\rightarrow {\rm s}_1) $ is
related to $R$ (37) as

\begin{equation}
	\pm S_{\rm min} ({\rm s}_2\rightarrow {\rm s}_1) =\lim_{T\rightarrow 0}[T\ln
	(R(T))],
\end{equation}

\noindent
 where $+,-$ correspond to ${\rm
	s}_2\stackrel{nf}{\rightarrow}1,2$ respectively.

In practice, however, there is always a lower limit for $T$ in
simulations, $T_l$, because the overall simulation time must not become
unrealistically long. That is why the use of (38) may, in practice,
introduce significant inaccuracy. To reduce the influence of the
pre-exponential factor we measure $R$ both at $T_l$ and at a slightly
higher temperature, $T_l+\Delta T$ ($T_l\gg\Delta T \stackrel{\sim}{>}
T_l^2/ S_{\rm min} ({\rm s}_2\rightarrow {\rm s}_1)$), so that:

\begin{equation}
	\pm S_{\rm min} ({\rm s}_2\rightarrow {\rm s}_1)\approx \frac{T_l^2}{\Delta T}
	\ln(\frac
	{R(T_l+\Delta T)}{R(T_l)}).
\end{equation}

\noindent The quantities on the left and right of Eq.\ (39) are shown
in Fig.7 respectively by the dotted line (theory) and by squares and
crosses (electronic and computer simulations respectively). The
agreement is satisfactory, given that $5 \stackrel{\sim}{<} S_{\rm
	min}/ T_l \stackrel{\sim}{<} 7$.

Note that the magnitude of the largest oscillation in action may
significantly exceed $U_{{\rm s}_2}-U_1$. This occurs if the initial
well--1 is adjacent to an external saddle ${\rm s}_2$ while its depth
is much less than that of the other well.

Finally, we comment on the experimental consequence of the cutoff of
the MPDTP, namely the drastic change of the time evolution of $J$ for
$t_f^{(1)} \stackrel{\sim}{<}t \ll t_f^{(2)}$: at $\Gamma<\Gamma_0$,
one may in principle make $T$ small enough that the sharp growth of
$J(t)$ at $t\stackrel{\sim}{<} t_f^{(1)} $ turns into a nearly constant
value at $ t_f^{(1)} \ll t \ll t_f^{(1)}
\alpha_{13}/(\alpha_{12}\alpha_{23})$ while, at $\Gamma>\Gamma_0$,
$J(t)\approx \alpha_{12}\alpha_{23}t$ over the whole relevant
time-scale: cf.\ the thin full and dashed lines in Fig.6(b). 

\section{IV. Roadmap for the subject}

In this section, we 
briefly
discuss potentially interesting directions of a development of the subject in future both science-wise and for applications,
in the subsections A and B respectively. 

\subsection{A. Scientific directions}

It is convenient to formulate open problems for the ranges $t\ll t_f^{(s)}$ and $ t_f^{(1)}\ll
t\stackrel{\sim}{<} t_f^{(2)}$ separately.

\subsubsection{1. Range of times being much less than time of the formation of guasi-equilibrium within a single/initial area of phase space}

It would be interesting to study the following issues.

\begin{enumerate}
	
	\item Details of the case
	considered above, including in particular an accurate study of: (i) oscillations of the
	exit velocity and $dS_{\rm min} /dt$ as the exit time goes, as well as (ii) the
	transition from a smooth $S_{\rm min} (t)$, with inflection points
	only, to an $S_{\rm min} (t)$ possessing folds. 
	
	\item Additional features
	characteristic of other types of 
	boundary or other types of transitions, in particular
	inter-well transitions in the  symmetric double-well potential - the case particularly relevant in the context of some promising application (see Sec. IV.B.1 below). 
	
	\item A careful
	consideration of the case with two absorbing walls while the initial
	coordinate is close to one of the walls and the initial velocity is directed towards the opposite wall, a case that is relevant e.g.\ to ionic
	channels \cite{hille,Zheng,Igor}. The preliminary analysis indicates
	oscillations of the flux in time. 
	
	\item Generalization for
	non-potential systems and/or non-white noise for which, unlike
	potential systems subject to white noise where switching between
	different MPEPs gives rise only to folds in $S_{\rm min} (t)$, we
	anticipate the possibility of jumps in $S_{\rm min} (t)$. Of a particular interest, the cases of various low-frequency noises \cite{low-frequency} and quasi-resonant noise \cite{Huang:19} are since they relate to many real systems.
	
	\item Pre-exponential factor.
	
	\item Multi-dimensional problems.
	
\end{enumerate}

\subsubsection{2. For the multi-equilibria case only: range of times in between the time-scale of the formation of quasi-equilibrium within the single well/area and that within all wells or, more generally, areas of attraction of all attractors}

\begin{enumerate}

\item The case with more than two barriers.

\item Pre-exponential factor.

\item Multi-dimensional problems.

\end{enumerate}

\subsection{B. Applications}

There may be various applications of the results described above. We restrict ourselves to a description of just two of them, which seem to us most promising.

\subsubsection{1. Measurements of noise intensity in a huge range}
 Thermometers or, more generally, meters of noise intensity, which we call further as {\it noisemeters} typically can measure temperature or noise intensity respectively in a quite limited range only.
In other words, the lower and upper limits of measurements are of the same order of magnitude or, at best, they differ by 1-2 orders of magnitude only. For example, a common room thermometer can measure temperature just in the range 280-320 K. 

Generally speaking, an estimate 
of temperature or noise intensity in case of a non-thermal noise (for the sake of brevity, we shall use one and the same notation $T$ for both cases) can also be based on a measurement of a quasi-stationary noise-induced escape flux. However, to the best of our knowledge, for real routine estimates of $T$ (i.e. for everyday or engineering purposes rather than for just scientific ones) it has not be used. Perhaps, the reason of this is the following: on the one hand, such a measurement is rather time-consuming and, on the other hand, the range of $T$ which can be thus measured is not very large. As for the lower limit, it is about $T_{qs}^{(l)}\approx \Delta U/12$, where $\Delta U\equiv U(q_{aw})-U(q_b)$ is the value of the \lq\lq barrier'' for a given value $q_{aw}$ (the lowest possible $\Delta U$ is limited with the lowest value of $q_{aw}-q_b$ which is possible to measure sufficiently accurately) while the denominator $12$ is explained by that, for larger values of the denominator, the escape probability is so low that it is impossible to measure the escape flux for a realistic time \cite{hoban}.
The main restriction for the estimate of $T$ by means of the measurement of the {\it quasi-stationary} escape flux relates to the upper limit of $T$ which can thus be measured: it is inherently limited from above 
with the value
$T_{qs}^{(up)}\approx \Delta U_{max}/3$, where $\Delta U_{max}\equiv U(q_s)-U(q_b)$ is the maximal possible value of the potential barrier (cf. Fig. 1). For larger values of $T$, the exponential (activation-like) factor in the dependence of the escape rate $\alpha_{qs}$ (2) on $T$ is not sufficiently sharp and the Kramers-Melnikov formula for  $\alpha_{qs}$ \cite{Melnikov:84,Melnikov_Meshkov:86,Melnikov:91} is not valid anymore. The restriction for the upper limit $T_{qs}^{(up)}$ is especially important in case of a non-thermal noise as such a noise may have a very large intensity and, moreover, its variation may be very large (constituting many orders of magnitude).  

Our results for the noise-induced escape at time-scales much less than the scale of the formation of the  quasi-equilibrium in a single well $t_{f}^{(s)}$ (7) (i.e. those described in Sec. III.A) in case of low friction, which is relevant first of all to nano/micro-mechanical resonators, promise to provide a possibility to measure noise intensity in the range varying by many orders of magnitude while using {\it one and the same} devise. Application-wise it may provide a great financial benefit.

There is no room here to provide details \cite{unpublished}. Rather we just give the main ideas and mention a few difficulties which may 
be encountered. We see two distinctly different options for an implementation of our ideas.

\begin{enumerate}

	\item {\it The setup with an \lq\lq absorbing'' wall}. As compared with the second option described below, the present setup allows us to immediately utilize the results presented in Sec. III.A (in particular, the explicit results for the parabolic approximation of the potential) and, besides, it might be favourable in terms of the duration of the required measurements and of the computation time required for the calculation of $T$ from the measurements. At the same time, the setup might give rise to a serious technical problem: each time when the system reaches the \lq\lq wall'' (being, in fact, just a given coordinate rather than a real wall), it should be somehow returned into the initial state (i.e. in the bottom of the well), and it is desirable for this transition to occur quickly, which may not be easily feasible. Generally speaking, such a return might be fulfilled by means of an interruption of an action of noise on the system (e.g., if noise acts due to an electric connection, then the corresponding connection may be switched off). Then, the time-scale for a single return is $
	\Gamma^{-1}$. If there is a possibility to strongly increase the diccipation, then the time-scale would further grately decrease. Another possibility is to introduce some additional action on the system which would transfer the system into the vicinity of the bottom of the well. We do realize that the problem of a fast return into the bottom of the well after reaching the coordinate $q_{aw}$ may not be trivial in reality, but meanwhile we assume that it can be resolved somehow. 
	
	Let us discuss the key points of the algorithm of the measurements.
		  Firstly, we should roughly estimate the quasi-stationary escape rate $\alpha_{qs}$. To this end, it would be sufficient to observe just a few (up to $10$) escapes and average them, thus obtaining the rough estimate for $\alpha_{qs}$. 
		  
		  If the relation $\alpha_{qs}\ll \Gamma$ holds true, then we may conclude that $T\ll \Delta U$  and therefore the Kramers formula for $\alpha_{qs}$ \cite{Melnikov:84,Melnikov_Meshkov:86,Melnikov:91,Kramers:40} may be readily used, so that we just need to gain more statistics in order to measure $\alpha_{qs}$ more accurately and then to calculate $T$ from it by means of the Kramers formula.
		  
		  Of the main interest in the present context is the complementary case: $\alpha_{qs}\gtrsim \Gamma$. It follows from this relation that $T\gtrsim \Delta U$. In this case, we should measure the flux at relatively small time-scales namely at $t\ll t_{esc}\equiv \alpha_{qs}^{-1}$. In order to obey this inequality while avoiding too poor statistics we should chose a compromise, namely to explore the time $t\approx t^{(N)}\equiv t_{esc}/N$ where $N$ is a moderately large number (about $5-6$). Then we need to compare $t^{(N)}$ with the step width, i.e. with $\pi/\omega_0$. If $t^{(N)}\gg\pi/\omega_0$, then the step structure is smeared at the time-scale $t^{(N)}$. It is convenient in this case to measure the escape flux at $t=t^{(N)}$ and $t=t^{(N+1)}$ and to compare with each other while T can be shown to obey the following formula:
		  
		  \begin{eqnarray}
		  	&&
		  	T=\frac{\Delta U}{\Gamma t_{esc}\ln\left(\frac{J(t^{(N)})}{J(t^{(N+1)})}\right)}, 
		  	\quad
		  			  	\frac{\pi}{\omega_0}\ll t^{(N)}
		  			  	\lesssim 
		  			  	\frac{1}{\Gamma},
		  	\\
		  	&&	  	
		  	t^{(K)}\equiv \frac{t_{esc}}{K},
\quad
t_{esc}\equiv \alpha_{qs}^{-1},
\quad
N \approx 6.
\nonumber
		  \end{eqnarray}
	  
	If $t^{(N)}$ is of the same order as $\pi/\omega_0$, then it may be 
	preferable
	to measure the flux at the centers of the first and second steps rather than at $t^{(N)}$ and $t^{(N+1)}$: the results are much less sensitive to an inaccuracy of a measurement of time. Using formulas in Eq. (19), 
	we obtain:

\begin{equation}
	   	T=\Delta U\frac{\omega_0/(2\pi \Gamma)}{\ln\left(J(t_{2})/J(t_{1})\right)}, 
	   	\quad
	   \frac{t_{esc}}{6} \sim\frac{\pi}{\omega_0} 	,
	\end{equation}
	where $t_n$ is defined in (19).
	
	Finally, if $t^{(N)}\ll\pi/\omega_0$, then one should use the function $S_{min}(t)$ calculated for $t=t^{(N)}$ and $t=t^{(N+1)}$ by methods described in Sec. III.A. Then $T$ can be calculated by means of the formula which is formally valid in a general form for any time-scale and for any position of $q_{aw}$:
	
	  \begin{equation}
		T=\frac{S_{min}(t^{(N+1)})-S_{min}(t^{(N)})}{\ln\left(\frac{J(t^{(N)})}{J(t^{(N+1)})}\right)}, 
	\end{equation}
	where  $t^{(K)}$ and $N$ are defined in Eq. (40) and we do not restrict the range of its validity to $t^{(N)}\ll\pi/\omega_0$ since Eq. (42) is valid in the much broader range: $t^{(N)}
	\lesssim
	\Gamma^{-1}$ 
	(Eq. (40) represents a partial case of (42) provided $q_{aw}$ lies sufficienntly close to the bottom of the well, so that the parabolic approximation of $U(q)$ works well)). Of course, various inaccuracies of experimental measurements and theoretical approximations put a limit for the lowest limit of the range of $t^{(N)}$ where Eq. (42) is valid and this determines the upper limit for values of $T$ which we can measure by means of such a method. The limitations will be discussed elsewhere.
	
	Even if to skip the range of very small times and to restrict ourselves to the range $t_{esc}/6\gtrsim \pi/\omega_0$, we can see from Eqs. (40) and (41) that our approach allows one to measure $T$ within the range characterized with the ratio of an upper and lower limits of the order of the quality factor $Q\equiv \omega_0/\Gamma$. A few more orders of magnitude may be added for the 
	account
	of a decrease of an effective $\Delta U$ (by means of shifting $q_{aw}$ closer to the bottom of the well). Quality factors of modern nano/micro-mechanical resonators can rather easily reach values $10^6-10^7$ \cite{Huang:19,dykman:22,Moser:13} and therefore our method provides a possibility to measure $T$ with one and the same device within a huge range of the order of $10^8-10^{10}$. 
	
	\item {\it The setup with transitions between bottoms of wells of a symmetric double-well potential}. If we use a device charecterized with a symmetric double-well potential (for example, it may be a buckled doubly-clamped beam \cite{Erbil:20} or a nanoparticle levitating in a bistable one-dimensional opticle trap \cite{Flajsmanova:20}) 
and consider transitions between close vicinities of bottoms of the potential wells, then the transitions in both directions are equivalent in the context of the transition probability and transition flux. Therefore there is no need to artificialy return the system into the initial state. 
As compared with the setup with the absorbing wall, this is a big advantage. One of disadvantages consists in that we cannot vary the magnitude of an effective barrier $\Delta U$. Besides, it is necessay to generalize the theory for this case, and the results certainly will not be expressed explicitly. On the other hand, the latter disadvantage (a necessity to use a complicated numerical procudure for a calculation $S_{min}(t)$) is not crucial: for a given potential $U(q)$, one will be able to calculate $S_{min}(t)$ once and forever, so that it will be used just as a known numerical function for any new measurement.

\end{enumerate}

We conclude this sub-section with the formulation of its main idea in an alternative form. We suggest to replace a straightforward measurement of temperature $T$ (or its equivalent in case of noise of a non-thermal origin) for a measurement of the escape/transition flux in an appropriate time range. The method allows to measure $T$ in a huge range using the following idea. When $T$ is of the order of or larger than an effective potential "barrier" in our system, then, knowing the theoretical dependence of the activation barrier on the time of a given noise-induced escape/transition, we can measure the relevant escape/transition time at which the activation barrier corresponds to the relevant flux. If a clock used in the time measurements exploites a periodic process with a very small period, then the clock can measure time in a huge range, being much larger than ranges in which a measurement of temperature by means of straightforward methods can be done. In a sense, {\it we suggest to reduce a measurement of temperature to a measurement of time while the latter can be measured in a much larger range than temperture can be conventionally measured}. 

\subsubsection{2. Accurate measurement of a damping parameter}

A linear damping parameter $\Gamma$ of an underdamped oscillator is typically measured as a half-width of a resonance curve in case when noise is absent (or negligible) while the driving amplitude is sufficiently small for a nonlinearity to play a negligible role for constrained vibrations (see e.g. \cite{Huang:19}). Such measurements are not very accurate however. Results reviewed in this paper might provide a method of a much more accurate measurement of a linear damping. We briefly describe it below.

We assume that we know temperature $T$ in the system and $U(q)$ in some vicinity to the bottom of the well with a high accuracy. Then we should {\it roughly} measure/estimate $\Gamma$ (which can be done with a few methods). As described in the previous item IV.B.1, we can readily define the value of $q_{aw}$  and choose time $t= t^{(N)}$ so that the corresponding action $S_{min}(t^{(N)})$ exceeds $T$ with the optimal factor $N=5-6$. Given that action is inversely proportional to $\Gamma$, the flux $J(t^{(N)})$ depends on $\Gamma$ in the activation-like manner i.e. very sharply. At the same time, the statisctics of escapes is not too poor (due to that $N=5-6$ is just {\it moderately} large), so that the flux can be measured with a high accuracy while the accuracy of the estimate of $\Gamma$ from the flux measurements is even larger with the factor of the order of $N$.

\section{Conclusions}

Our paper reviews results on noise-induced escapes and transitions at time-scales preceding the formation of equilibrium/quasi-equilibrium for the case of white noise and linear damping with a small or moderate value of the damping parameter $\Gamma$, and discussess interesting open problems 
as well as a couple of promising applications.

The escapes/transitions at small time-scales occur quite qifferently from those at time-scales exceeding the time-scale of the equilibrium/quasi-equilibrium formation $t_{qe}\sim \Gamma^{-1}$, and the corresponding probability flux is exponentially smaller. The strongest difference concerns the case of small damping ($\Gamma\ll\omega_0$, where $\omega_0$ is a frequency of weak eigenoscillations of the system), and the smaller the escape/transition time $t$ is, the larger the difference is. Rougly speaking, the activation energy $S_{min}$ for the escape/transition at time $t$ is of the order of $\Delta U/(\Gamma t)$, where $\Delta U$ is a relevant potential barrier or a difference of relevant energies. Thus, for the relevant range $t\ll\Gamma^{-1}$, the activation energy $S_{min}(t\ll\Gamma^{-1})$ greatly exceeds the conventional activation energy $\Delta U$ for the quasi-stationary flux. As $t$ decreases, a topology of the most probable escape path (MPEP) changes: a number of turning points decreases at such values of $t$. Sometimes, this results in a continuous change of the MPEP and, sometimes, in a jump-wise change. The most interesting feature of the function $S_{min}(t)$ is its step-wise form, which is due to the aforementioned change of the MPEP topology. The time-scale of the step is $\sim\pi/\omega_0$, and steps are most distinct at times $t$ of the order of the period of the eigenoscillation.

For potentials with more than one barrier, the direct escape flux over both barriers is characterized with the activation energy $S_{dir}$ which, as function of $\Gamma$, undergoes oscillations in the range small values of $\Gamma$ and a large increase in the range of moderate values (typically limited from above by the value $2\omega_0$). These variations are related to variations of the MPEP, including in particular the variation of its topology responsible for the oscillations in $S_{dir}(\Gamma)$.

We also briefly discuss various open related problems which seem to us interesting and, in somewhat larger detail a couple of potential applications, namely: (i) a measurement of temperature or of noise intensity of a non-thermal noise in a very broad range, (ii) a rather accurate measurement of the damping parameter.

\section{ACKNOWLEDGEMENTS}
We acknowledge discussions with Mark Dykman, Yuri Rubo and Yuri Semenov.


\begin{thebibliography}{100}
\bibitem{footb} The absorbing wall is equivalent to the barrier falling
   down to $-\infty$, which is illustrated by Fig. 6(a).
\bibitem{footc} The choice of $t_l$ is quite arbitrary: it should
   exceed the relevant time-scale and is introduced purely to reduce the
   overall time of experiment.
\bibitem {formation} If the wall is close to the bottom, the estimate
   (7) can be obtained by consideration of the parabolic potential: the
   Fokker-Plank equation is then solved explicitly \cite {wang_uhlenbeck}.
   It follows directly that, if the initial state corresponds to the
   bottom of the well i.e.\ $W(q,\dot{q},t=0)
   =\delta(q-q_b)\delta(\dot{q})$, then $W(q=\sqrt{2\Delta U}/\omega_0,\dot{q}=0,t)$ 
   evolves close to the
   equilibrium value for the time described by Eq.(7). In another
   characteristic case, when the wall is close to the saddle (cf.\
   $q_{aw}^{(3)}$ in Fig.1), the estimate (7) is also typically valid: it
   agrees with the explicit results \cite{schneidman} for the underdamped
   and overdamped limits provided that the curvatures at the saddle and at
   the bottom are of the same order, which 
   is typically the case:
   cf.\ $U(q)$ (1).
\bibitem {infinity} If $n\rightarrow \infty$, so that $t_n\rightarrow
   \infty$, then $S(t_n)$ reduces to $\Delta U_p$ while the MPEP reduces
   to the time-reversal of the noise-free relaxational trajectory from the
   state $\{q=q_{aw},\dot{q}=0\}$, thus recovering the conventional
   Arrhenius activation energy for the quasi-stationary flux 
   (Refs. 1-4)
   and the corresponding MPEP 
   (Refs. 34 and 31),
   respectively. The reduction of the asymptotic (at $t\rightarrow
   \infty$) activation energy and MPEP to their conventional forms is
   valid for an arbitrary $U(q)$.
\bibitem{footf} To avoid the singularities encountered by
   $\Gamma^{\prime}$ at the turning points, one may use the variable
   $G\equiv \Gamma^{\prime}\dot{q}$ instead.
\bibitem{footg} A similar equation was obtained in 
   Ref. 34
   but in a
   different context and by a different method.
\bibitem{footh} The numerical search is more difficult than in the case
   of a single solution.
\bibitem{footnontherm} Note that the additional noise source does not
   need necessarily be of a thermal origin. In SQUIDs, for example, it
   could be an external magnetic flux noise.
\bibitem {additional} Our qualitative analysis indicates that, in
   addition, some new features may appear in the multi-well case, but this
   has not yet been studied in detail.
\bibitem{footi} For the sake of brevity, we refer to region 3 as an
   \lq\lq attractor" too.
\bibitem{footj} Some of the results \cite {soskin} are closely related
   to some results of \cite{gt} which were obtained by a different method
   and in a different context.
\bibitem {others} In \lq\lq long'' jumps in a multi-barrier potential
   with equal or nearly equal barriers, a monotonic
   exponential dependence on friction was found in earlier works (see e.g. \cite{new} and \cite{Melnikov:91} respectively). However,
   the obvious reason that oscillations and cusps were not found in those
   studies is that they can only arise in the case of distinctly different
   barriers levels.
\bibitem{footk} Note however that pre-exponential time-scales in other
   systems (e.g.~in SQUIDs or NEMS/MEMS6) may be much smaller, so that the relevant flux
   may be readily simulated and/or measured.
\end{thebibliography}
\end{document}